\documentclass[journal]{IEEEtran}
\usepackage{amsmath,amssymb,epsfig,xcolor} 
\usepackage{tabularx,multirow,array}
\usepackage{times}
\usepackage{booktabs}
\usepackage{multirow}
\usepackage{soul}
\usepackage{enumerate}
\usepackage[countmax]{subfloat}
\usepackage{algpseudocode}
\usepackage{algorithm}
\usepackage{subfigure}
\usepackage{threeparttable}
\usepackage{pifont}
\usepackage{cite}
\usepackage{url}

\newcommand{\xmark}{\ding{55}}%

\newtheorem{theorem}{Theorem}

\begin{document}

\title{Computing Exact and Approximate Blocking Probabilities in Elastic Optical Networks}

\author{{Sandeep Kumar~Singh, ~\IEEEmembership{Student Member,~IEEE} and Admela~Jukan,~\IEEEmembership{Member,~IEEE}}
\thanks{The authors are with the Department of Electrical and Computer Engineering, Technische Universit\"at Carolo-Wilhelmina zu Braunschweig, Braunschweig 38106, Germany (e-mail: \{sandeep.singh, a.jukan\}@tu-bs.de). This paper is under submission, and uploaded here only for comments and suggestions, and not for any commercial use.}
}

\maketitle

\begin{abstract}
In this paper, we propose the first exact Markov model  for connection blocking analysis in elastic optical networks, based on the occupancy status of spectrum slices on all links due to arrivals and departures of various classes of connections in a network. Since the complexity of the exact Markov model grows exponentially with the link capacity, number of links, routes, and classes of demands, we further advance the state-of-the-art in computing approximate blocking probability in elastic optical networks and propose two novel approximations, i.e.,  load-independent and load-dependent. These approximations are used to compute state-dependent per-class connection setup rates in multi-class elastic optical networks with or without spectrum converters by taking into account the spectrum fragmentation factor in each state.  We validate approximation analysis by exact  and/or simulation results, and show that load-independent  and load-dependent approximations can be more accurately used than previously proposed approximations, under a random-fit (RF) and a first-fit (FF) spectrum allocation policies. The approximate results match closely with the exact model, for smaller networks, and with the simulations under a variety of network scenarios. 
\end{abstract}


\section{Introduction}
\par \IEEEPARstart{F}{lexi}-grid elastic optical networks (EONs) divide optical spectrum into units of spectrum grids (slices), that can be flexibly allocated in form of ``just-enough" spectrum amounts to variable bandwidth demands \cite{jinno2009spectrum}. Elastic spectrum allocation thus generally increases spectrum utilization in comparison to fixed spectrum systems. At the same time, connection requests in these networks can also be blocked due to the fragmentation of spectrum occupancy, in addition to resource unavailability. In fragmentation states, even though sufficient, but scattered, spectrum maybe available in the network, a connection request maybe blocked if there is no required number of continuous and contiguous slices available for a new bandwidth demand. The spectrum continuity and spectrum contiguity are in fact two fundamental constraints for routing and spectrum allocation in elastic optical networks. The \emph{spectrum continuity constraint} requires an incoming connection (lightpath) request  to be provisioned all-optically over the same set of subcarrier slices in all links it traverses. The constraint called \emph{spectrum contiguity constraint} means that a connection request demanding multiple subcarriers needs to be allocated over adjacent frequency slices. These two constraints along with the resulting spectrum fragmentation have been subject to much research in optimal routing and spectrum allocation (RSA) schemes, presenting an NP hard problem \cite{wang2011study}, for which the exact blocking model has not been formulated yet. 

\par In this paper, we address this grand challenge for the first time by proposing an exact Markov model for computing connection blocking in EONs, where a network state is represented by occupancy of individual spectrum slices  on all network links. Since the complexity of the exact Markov models grows exponentially with the link capacity, number of links, routes, and classes of demands, we further advance the state-of-the-art in computing approximate blocking probability in elastic optical networks. Note that implementing an approximation that is tractable, yet sufficiently accurate has been studied in the past, and was shown not to be a trivial task either. It is also analytically hard to taking into account the effect of fragmentation while deriving the connection setup rate, i.e., the effective arrival rate at which a link allows connections to be setup in a given spectrum occupancy state. The reason is that a given number of occupied slices could be represented by fragmented as well as non-fragmented spectrum patterns, and the fragmented states could not accept an incoming connection request. The estimation of the connection setup rates taking into account the fraction of time a link stays in non-blocking states can be obtained by monitoring the link state occupancy over a long period of time, as shown by simulations by Reyes \textit{et al.} \cite{reyes2016reward}. In absence of such monitoring information, the progress towards a tractable and close to accurate approximate blocking analysis has been challenging, and slow. To address these challenges, we develop two novel approximations. Finally, this paper compares the novel approximate blocking performance with the exact and/or simulation results to evaluate the approximation analysis. The approximate results are accurate under a range of scenarios, including varying link capacities, classes of demands, traffic loads and network topology.

\par The rest of this paper is organized as follows. Section \ref{sec:relatedWork} presents the related work and our contribution. We present the exact network blocking model in Section \ref{sec:ExactModel}.  In Section \ref{sec:stateDescription}, we describe a novel reduced state model, and identify non-blocking and blocking exacts states for each occupancy state in the reduced state model. We present model assumptions and approximation approaches towards calculating the probability of acceptance of a connection in Section \ref{sec:stateDescription}. Section \ref{sec:Methodology4BP} presents a multirate loss model to computing  connection setup rate, departure rate and approximate blocking probability in EONs. We evaluate the performance in Section \ref{sec:evaluation}, and conclude the paper in Section \ref{sec:conclusion}. 

\begin{table*}[t]
\centering
\scriptsize
\caption{Comparison of related work and this work w.r.t. various factors used in BP computation in EONs without SC.}
\label{table:RelatedWork}
\begin{threeparttable}
\begin{tabular}{|c|c|c|c|c|c|c|c|}
\hline
BP         & work              & Approach &  Exact states  & Contiguity & Continuity & Reduced-load app. & Link-load correlation  \\ \hline \hline
\multirow{2}{*}{Exact}                    &  \cite{beyranvand2014analytical}                 &  CTMC     & Partially    & \checkmark   &  \checkmark  &  NA  &  \xmark    \\ \cline{2-8}
                          &  This paper                 &  CTMC     & \checkmark    & \checkmark   &  \checkmark  &  NA  &  \checkmark     \\ \hline
\multirow{6}{*}{App.}         & \multirow{2}{*}{\cite{beyranvand2014analytical}} &   Kaufman  \cite{kaufman1981blocking}     & \xmark  &  \xmark   & \xmark   &  \xmark &  \xmark  \\ \cline{3-8} 
                                              &                   &  Binomial \cite{peng2013theoretical}        &  \xmark  & Partially    & Partially   &   \xmark  & \xmark    \\ \cline{2-8}                                              
                                            &                    &     EES  \cite{kuppuswamy2009analytic}    & Partially   & \checkmark     &  Partially  &  \checkmark   & \xmark  \\ \cline{3-8} 
                                            &   This  paper       &  SOC  \cite{kuppuswamy2009analytic}      & Partially    &  \checkmark    &  Partially  & \checkmark  & Partially   \\ \cline{3-8} 
                                            &                   &  Uniform   \cite{kuppuswamy2009analytic}         &  \xmark    &  Partially    & Partially    &  \checkmark & \xmark     \\  \hline
\end{tabular}
\begin{tablenotes}
   \item App.: Approximation; NA: Not applicable; EES: Equiprobable exact states; SOC: Slice occupancy correlation.
   \end{tablenotes}
\end{threeparttable}
\end{table*}

\section{Related Work and Our Contribution}\label{sec:relatedWork}
\par The connection blocking analysis in elastic optical networks has been studied in a few notable works, including \cite{yu2013first,peng2013theoretical,beyranvand2014analytical}. Most of the work previously reported, starts with an analysis of blocking in a single optical link, -- which is due to complexity, building it up to a computation of blocking for a network. In \cite{yu2013first}, an exact blocking probability of a single link (with a small scale capacity in number of spectrum slices) was analyzed by modeling the bandwidth occupancy as continuous-time Markov chain (CTMC) under a random-fit (RF), which randomly allocates spectrum to connection requests, and a first-fit (FF), which allocates first available slices in an ordered set of slices, spectrum allocation methods. In our previous work \cite{singh2017analytical}, we used that same exact CTMC model with additional reconfiguration states to analyze blocking in a link with a reactive and a proactive connection reconfiguration methods. The network blocking analysis by a so-called exact solution was given by Beyranvand \emph{et al.} \cite{beyranvand2014analytical}, where network states are defined by a spectrum union operation of exact link states on a multi-hop route, without differentiating among spectrum patterns formed by overlapping routes and without consideration of link-load dependency. Therefore, the presented exact results for a 2-hop network do not generally match simulation results.  

\par Because the computational complexity of an exact link model increases exponentially with number of spectrum slices, \cite{peng2013theoretical} and \cite{beyranvand2014analytical} presented approximation models.  In \cite{peng2013theoretical}, authors used a binomial distribution approach to compute the probability of a required number of free consecutive slices on a link by assuming the average carried load per slice as the slice occupancy probability. The same approach was used in \cite{beyranvand2014analytical} to approximate an idle slice probability using link state probabilities, which is obtained by a Kaufman's formula \cite{kaufman1981blocking} without considering the spectrum fragmentation caused by bandwidth demands and the RSA constraints.  More in detail, the Kaufman blocking probability solutions in \cite{beyranvand2014analytical} were shown to match the exact analysis, and  simulation results, for cases where RSA constraints are relaxed. The binomial approach \cite{peng2013theoretical} was shown useful for cases with a relatively small scale link with capacity in number of spectrum slices, since it tries to estimate the availability of contiguous and continuous free slices on a route. It should be noted that both Kaufman and binomial approaches do not consider valid spectrum patterns (exact states) and that neither on a single link nor on a multi-hop route. Thus, probability of finding a required number of contiguous and continuous free slices would be very inaccurate in these two approaches. Recently, an important step towards reducing an exact link-state description model to a Macrostate model (states are denoted by connections per class) was shown by Reyes \textit{et al.} \cite{reyes2016reward}. They estimated the connection setup rates in non-blocking states using a link-simulation approach, and used it for controlling the  call admission in EONs. 
 
To advance the previous studies, we introduce for the first time an \emph{exact Markov model} for a network, wherein all possible network states and transition among them are defined to obtain the network state probabilities and connection blocking. Following which, due to scalability issue of the exact network model, we propose a  \emph{reduced state Markov model}, wherein link states are represented by total occupied slices, and  the connection setup rates in link states consider spectrum fragmentation factor to compute approximate blocking probabilities in EONs, with and without spectrum conversion, using  load-independent and load-dependent approximations. Additionally, we consider a \emph{reduced-load approximation}, which helps in calculating the effective load of the combined connections on a link, and an \emph{independence link} assumption, i.e.,  spectrum occupancies of links are statistically independent, using a multirate loss model. We note that the multirate loss model was originally developed for fix-grid Wavelength Division Multiplexed (WDM) networks \cite{birman1996computing,kuppuswamy2009analytic} to  compute approximate blocking probability. The multirate loss model in \cite{birman1996computing,kuppuswamy2009analytic} uses a request acceptance probability term to compute connection setup rates and blocking probability. However, without using the exact network model, the exact calculation of the probability that a bandwidth request is accepted on a route in an elastic optical network, which also needs to consider contiguity constraint, is still an unsolved problem. 

Thus, we propose two different load-independent approximations: (i) Uniform, and (ii) Equiprobable Exact States (EES). While the Uniform approximation assumes that a link occupancy is uniformly distributed over slices without paying attention to valid exact states, the EES approach assumes that observing a link occupancy (denoted by $x$)  among exact link states that have same number of occupied slices ($x$) is equiprobable. Notice that these approximations do not handle slice occupancy correlation (SOC) among fragmented and non-fragmented exact states differently, which is important especially in scenarios with low network load, but also in the FF spectrum allocation policy. Therefore, we propose a load-dependent SOC approximation to calculate the probability of acceptance of a request in EONs considering the fragmentation and average link occupancy for a given network load. 

To illustrate how the proposed models advance the state-of-the-art, Table \ref{table:RelatedWork} summaries the main factors of our analysis and the related work in computing exact and approximate blocking probabilities in EONs, all without spectrum conversion (SC). (Spectrum conversion means that if, for instance, two contiguous slices $s_1$ and $s_2$ are used on link 1, they can be \emph{converted} in an intermediate optical node and allocated to contiguous slices $s_3$ and $s_4$ on subsequent link 2.  A comparison of different approaches for blocking probability computation in EONs with SC can be given by omitting the continuity constraint in Table \ref{table:RelatedWork}.) It should be finally noted that no approximations, including ours, consider exact network states, as shown in Table \ref{table:RelatedWork}. However, the EES and SOC approaches consider exact link states only for the RF scenario in a single link system. Furthermore, the contiguity and continuity constraints are shown only partially true for methods in Table \ref{table:RelatedWork} if a method does not consider valid spectrum patterns (exact states) while computing the probability of required free contiguous and continuous slices on a route. 

\par While the statistical link independence assumption is common and used by all approximations, including this paper, for computing blocking probability in EONs, it is strong and critical, as it ignores the load correlation factor among  links. Although we do consider load correlation among slices on each individual link in the SOC approximation, the load-correlation among links are not considered due to complexity and scalability issues. At the same time, nonetheless, the approximate blocking results that we obtain in this paper for different operation modes (based on policies, with/without SC)  are promising, as they match closely to the exact, or simulation results for the RF and FF policies for most of the traffic loads and classes of demands. Notably, we use a CTMC model and the multirate loss model \cite{birman1996computing,kuppuswamy2009analytic} to derive exact and approximate blocking probabilities, respectively, under the following operation modes:  RF policy without SC (RF), FF policy without SC (FF), RF policy with SC (RF-SC), and FF policy with SC (FF-SC). 
It should be noted that  all operation modes assume that the spectral contiguity constraint must be satisfied while admitting a request, otherwise it simplifies to WDM scenarios which have been well investigated in the past \cite{birman1996computing,kuppuswamy2009analytic}. Also notice that the scenarios with SC make sense only in multi-hop routes in EONs, as SC helps in relaxing the continuity constraint. 

\begin{table}[t]
\caption{Notations and the parameters used in the models}
 \centering
\scriptsize
  \begin{tabular}{@{}ll@{}}
\toprule
{\bf Notation} & {\bf Description} \\ \midrule
     $ C  $      &   Total number of spectrum slices (or capacity units) per link            \\ \midrule
     $ K  $      &   Number of connection classes. Note that classes $k = 1, 2, \cdots, K$              \\ \midrule
     $ \lambda_k^o $      &    Arrival rate of class $k$ connections of an OD pair $o \in \mathcal{O}$                \\ \midrule
     $\mu_k$    & Service rate of a class $k$ connection; mean service time $t_c=1/\mu_k$ \\ \midrule
     $ \textbf{d}$         & $ \equiv (d_1, d_2, \dots, d_K)$, where $d_k$: bandwidth (in slices) of class $k$                    \\ \midrule
     $r(o)$          & Route of a path request $o \in \mathcal{O}$ consisting of some links $j \in \mathcal{J}$   \\ \midrule
     $\textbf{s}$      &   $ \equiv (s_1, s_2, \dots, s_C)$,  where $s_c$: free or occupied state of an $c^{th}$ slice  \\ \midrule
     $\textbf{n}$      &   $ \equiv (\textbf{n}^1, \textbf{n}^2, \dots, \textbf{n}^\mathrm{r})$,  where $\textbf{n}^o$ is the set of connections $\in o$.      \\ \midrule
     $\textbf{n}^o$      &   $ \equiv (n_1^o, n_2^o, \dots, n_K^o)$,  where $n_k^o$ is the \# of class $k$ connections $\in o$.      \\ \midrule
     $n_k^o(\textbf{n})$   & Number of class $k$  connections of an OD pair $o$ in $\textbf{n}$   \\ \midrule  
    $n_k^o(V_i)$   & Number of class $k$  connections of an OD pair $o$ in a network state $V_i$   \\ \midrule
     $\Gamma_{V_i}^{o,k+}$ &   Set of possible states after a class-$k$ request arrives on route $r(o)$ in $V_i$  \\ \midrule
     $\Gamma_{V_i}^{o,k-}$ &   Set of possible states after a class-$k$ connection on $r(o)$ departs from $V_i$  \\ \midrule
     $f_m(\textbf{s}_i)$ & Size of the largest block of consecutive free slices in a link state $\textbf{s}_i$ \\  \midrule
     $\Omega_S(x)$          & Set of exact link states $\textbf{s}_i$ representing total occupancy of $x$ slices    \\ \midrule  
     $\mathbb{NB}(x, k)$ & Set of non-blocking exact states with occupancy $x$ for class $k$ requests \\ \midrule
     $\mathbb{FB}(x, k)$ & Set of fragmentation blocking exact states with occupancy $x$ for class $k$ \\ \midrule
     $\mathbb{RB}(x, k)$ & Set of resource-blocking exact states with occupancy $x$ for class $k$ \\ 
\bottomrule
\end{tabular}
\label{table:Notations}
\end{table}

\begin{figure*}[t]
 \centering
\includegraphics[width=0.95\textwidth]{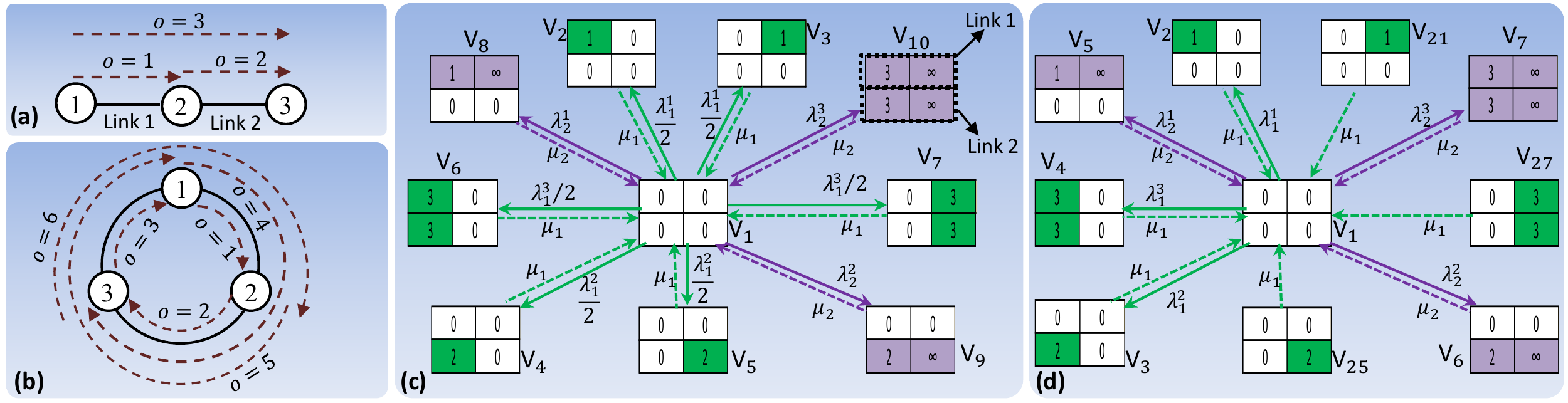}
\vspace{-0.3 cm}
  \caption{A 2-link topology with 3 OD routes, and a 3-node ring with 6 OD routes are shown in (a) and (b), respectively. State transitions from and into an empty network state ($V_1$) in a 2-link network with 2 slices per link  are shown due to arrivals and departures of connections on 3 OD routes with  two classes of demands $d_k = \{1, 2\}$ slices under the RF (c) and the FF (d) scenarios without SC, where each state is shown with occupancy of link 1 (2)  at top (below).}
 \vspace{-0.3cm}
\label{fig:NetworkStates}
\end{figure*}
\section{Exact Blocking Analysis} \label{sec:ExactModel}
In the Section, we present an exact CTMC model for computing exact blocking in EONs. The notations and definitions of some of the parameters used in the model are listed in Table \ref{table:Notations}. Let us list below all assumptions for the CTMC model for computing exact blocking probability in an arbitrary EON topology with $N$ nodes, $J$ unidirectional fiber links (belongs to set $\mathcal{J}$), and $C$ spectrum slices per link.

 \begin{itemize}
 \item Arrivals of class $k \in \{1, 2, \cdots, K\}$ connection path requests between an origin-destination (OD) node-pair $o (\in\mathcal{O})$  follow Poisson process with arrival rate $\lambda_k^o$ , and connection holding (service) time is exponentially distributed with mean $1/\mu_k$. We assume that the arrivals and departures are statistically independent.
 \item Each OD pair path request $o (\in\mathcal{O})$ is routed on a predetermined shortest path $r(o)$, and spectrum is allocated according to a given scenario: RF, RF-SC, FF, or FF-SC. 
 \item An OD pair request $o$ with bandwidth demand of $d_k$ slices is accepted in an EON  iff there are sufficient ($\geq d_k$) contiguous and continuous free slices  on its predetermined route $r(o)$. However, when network nodes are  equipped with spectrum converters, then the continuity constraint does not need to be satisfied.
\end{itemize} 

\subsection{Generation of Exact States, and State Transitions}
To compute exact blocking in EONs, we need to first define the states of a Markov chain, which are created by the allocation and deallocation of spectrum to  various classes of connections between different origin-destination (OD) node-pairs. 
Let us represent a free slice by 0, and an occupied slice by either $o$ or $\infty$ depending on whether the occupied slice is the start or the remaining bandwidth occupancy of a class $k$ connection on route $r(o)$. For example,  an empty network state $V_1$ of a 2-link EON, shown in Fig. \ref{fig:NetworkStates}(a), with 2 slices fiber link without any connection is represented by $V_1 = (\textbf{s}_1^1,\textbf{s}_1^2) = \{(s_1^1,s_2^1),(s_1^2,s_2^2)\}=\{(0,0),(0,0)\}$, where $\textbf{s}_i^j$ represents the $i^{th}$ network state occupancy on a link $j, j=1,2$, and $s_c^j$ shows the free or occupied status of an ordered (left to right) slice $c, c=1,2$ on a link $j$. Now, to illustrate the formation of a few other states, let us assume that  a new class 1 connection request arrives on an OD pair $o=1$  in an empty state $V_1$ with a bandwidth demand $d_1=1$ slice in the 2-link network with link capacity $C=2$ slices. Then, the spectrum can be allocated in one of two different ways under the RF policy, as shown by network states $V_2$ and $V_3$ in Fig. \ref{fig:NetworkStates}(c), where the top and bottom link states represent spectrum occupancy on links 1 and 2, respectively. Similarly, an arrival of a class 2 request with demand $d_2=2$ consecutive slices on an OD pair $o=2$ ($o=3$) in state $V_1$ will cause the network state to transit to a state $V_9$ ($V_{10}$). On the other hand, under the FF spectrum allocation policy, a new arrival ($d_1=1$ slice) on an OD pair $o=1$ in an empty state $V_1$ will trigger the network transition to only one network state  $V_2=\{(1,0),(0,0)\}$, where the first slice is allocated on link 1 in Fig. \ref{fig:NetworkStates}(d). Note that an exact network state space $\Omega_V$ for a given set of routes, classes of demands and link capacity vary based on the operation modes $M$:= RF, FF, RF-SC, FF-SC. Algorithm \ref{algo:ExactNetStates} describes a way to create valid network spectrum patterns (exact states), identifying transitions among them using a function $A(\cdot)$ and  blocking states using $B(\cdot)$, which is explained below.  

\begin{algorithm}[ht!]
\caption{Find $\Omega_V, A_{|\Omega_V| \times |\Omega_V| \times| \mathcal{O}|\times K},B_{|\Omega_V| \times |\mathcal{O}| \times K}$}
\begin{algorithmic}[1]
\State Given: $M, C, |\mathcal{J}|, \textbf{d}=\{d_1, d_2, \ldots,d_K\}, \forall o: r(o)$. 
\State Initialize $\Omega_V \leftarrow V_1:= zeros_{|\mathcal{J}|\times C} ; i \leftarrow 1; A, B\leftarrow \emptyset$.
\Repeat  
          \For {$(o, k)=(1,1): (|\mathcal{O}|, K)$} // \textit{due to allocation} 
    		  \State  \parbox[t]{\dimexpr\linewidth-\algorithmicindent}{$\Gamma_{V_i}^{o,k+}\!\!\! \leftarrow$States after possible allocation of $d_k$ slices \\ on $r(o)$ in a state $V_i$ based on operation mode $M$}
    		  \State $B(i,o,k) \leftarrow 1$, if $\Gamma_{V_i}^{o,k+}=\emptyset$;  otherwise  0 
    		  \For {each state $L_j \text{ in }  \Gamma_{V_i}^{o,k+}$} 
    		        \If {$L_j \in \Omega_V$}
    		           \State  $t \leftarrow \text{index of } L_j \text{in } \Omega_V, A(i,t,o,k)\leftarrow +k$ 
                    \Else    		           
    		           \State  $\Omega_V \leftarrow [\Omega_V,L_j], A(i,|\Omega_V|,o,k)\leftarrow +k$
    		        \EndIf 
    		  \EndFor    
    	 \EndFor
          \For {$(o, k)=(1,1): (|\mathcal{O}|, K)$}  // \textit{due to Deallocation}
    		  \State  \parbox[t]{\dimexpr\linewidth-\algorithmicindent}{$\Gamma_{V_i}^{o,k-}\!\!\!\leftarrow$States after deallocation of connection(s) \\ from $V_i$ with bandwidth of $d_k$ slices on route $r(o)$}
    		   \For {each state $L_j \text{ in }  \Gamma_{V_i}^{o,k-}$}      
    		        \If {$L_j \in \Omega_V$}
    		           \State  $t \leftarrow \text{index of } L_j \text{in } \Omega_V, A(i,t,o,k)\leftarrow -k$ 
                    \Else    		           
    		           \State  $\Omega_V \leftarrow [\Omega_V,L_j], A(i,|\Omega_V|,o,k)\leftarrow -k$
    		        \EndIf 
    		  \EndFor       
    	 \EndFor    
\State $i \leftarrow i+1$\Until{$i > |\Omega_V|$}   
\end{algorithmic}
\label{algo:ExactNetStates}
\end{algorithm}

A network state in the exact Markov chain is represented by a $|\mathcal{J}|\times C$ matrix, where an element $(j,c)$ represents the status of a $c^{th}$ slice  on link $j$, where $j=1,\ldots, |\mathcal{J}|; c=1,\ldots,C$. Algorithm \ref{algo:ExactNetStates}  initializes a network state in an empty state $V_1$,  i.e., $(j,c)=0, \forall j, \forall c$. Next, in Steps 4--14 for each combination of routes and classes, i.e., $(o,k)$, a state $V_i$ tries to allocate a class $k$ demand on route $r(o)$ while satisfying the RSA constraint(s) for a given operation mode M (e.g., RF, FF, RF-SC, FF-SC)\footnote{Note that the RF-SC (FF-SC) tries to first allocate a multi-hop connection $(o,k)$ over a random (first) set of contiguous and continuous free slices, and when the required free slices are not aligned over the route $r(o)$ then the continuity constraint is relaxed for the allocation. The reason is that an SC operation should help in reducing blocking, and we observe that while allocating spectrum in a state that has sufficient required continuous and continuous free slices, if non-aligned contiguous slices are selected then blocking may become higher (especially in RF-SC) than without SC scenario.}. A set of possible states due to arrival of a connection in the state $V_i$ is stored in a set $\Gamma_{V_i}^{o,k+}$ in Step 5, and states in $\Gamma_{V_i}^{o,k+}$ which are not present in the network state space $\Omega_V$ are appended to $\Omega_V$ in Step 11. Additionally, a function $A(\cdot)$ is updated to track the transition among states. Furthermore, if required free slices can not be allocated to an arrival ($o,k$) (Step 6) then the state $V_i$ is identified as a blocking state for the arrival  $(o,k)$ by setting $B(i,o,k)=1$. 
New network states are also created due to departures of connections (in FF, FF-SC and RF-SC), so Steps 15--24 try to capture new states due to deallocation of spectrum, and also updates the transitions among states. The (de)allocation process in each state $V_i, i\geqslant 1$ is checked until no more new network states are created. Thus, the network state space $\Omega_V$ converges, and the total number of network states is $|\Omega_V|$.

\subsection{Exact State Probabilities and Blocking Analysis}
\par After generating all exact states and transitions among them, the global balance equation (GBE) of a network state $V_i, i =1, \cdots, |\Omega_V|$ can be obtained by 
\begin{multline}
\left(\sum_{(o,k)=(1,1), \Gamma_{V_i}^{o,k+}\neq \emptyset}^{(|\mathcal{O}|, K)}\!\!\!\!\!\!\!\!\!\!\!\!\!\!\!\! \lambda_k^o  \,\,\,\,\,\,\,\,\,\,\,\, + \sum_{(o,k)=(1,1)}^{(|\mathcal{O}|, K)}  n_k^o (V_i) \mu_k  \right ) \pi(V_i) = \\
\sum_{t=1, t\neq i}^{|\Omega_V|} \left(\sum_{(o,k)=(1,1), V_i \in \Gamma_{V_t}^{o,k+}}^{(|\mathcal{O}|, K)} \!\!\!\!\!\!\!\!\!\!\!\!\!\!\!\!\!  \lambda_k^o/|\Gamma_{V_t}^{o,k+}| \,\,\,\,\, + \!\!\!\!\!\!\!\!\!\! \sum_{(o,k)=(1,1), V_i \in \Gamma_{V_t}^{o,k-}}^{(|\mathcal{O}|, K)} \!\!\!\!\!\!\!\! \mu_k\right) \pi(V_t) 
\label{eqn:GBEnetwork}
\end{multline}
where, left hand side (LHS)  represents the output flow rate from the state $V_i$ having steady state probability $\pi(V_i)$, while the right hand side represents input flow rate into the state $V_i$. More precisely, the first (second) term in LHS  of Eq. \eqref{eqn:GBEnetwork} represents the output rate due to arrivals (departures) in (from) $V_i$, and the first (second) term in RHS is due to arrivals (departures) in (from) other states $V_t$ that lead to the state $V_i$.  
As an example, under the RF scenario, the GBE for a state $V_1$  in Fig. \ref{fig:NetworkStates}(c) is given by $(\sum_{o=1,k=1}^{o=3,k=2}\lambda_k^o)\pi(V_1) =   \mu_1(\sum_{t=2}^7 \pi(V_t)) + \mu_2(\sum_{t=8}^{10}\pi(V_t))$. Notice that transitions from and into the state $V_1$ occur due to arrivals in $V_1$ and departures from other states, respectively. However, for example, the GBE of a state $V_2$ would also include rate $\lambda_1^1/2$ ($\mu_1$) in its RHS (LHS) due to an arrival (departure) of class 1 connection on OD pair route $1\rightarrow 2 (\text{ i.e., } o=1)$ in (from) the state $V_1$ ($V_2$).  Similarly, the GBE of state $V_1$ under the FF scenario in Fig. \ref{fig:NetworkStates}(d) can be obtain as $(\sum_{o=1,k=1}^{o=3,k=2}\lambda_k^o)\pi(V_1) =   \mu_1(\sum_{t=2}^4 \pi(V_t)+\pi(V_{21})+\pi(V_{25})+\pi(V_{27})) + \mu_2(\sum_{t=5}^7\pi(V_t))$, which allocates only first available free slices. Under the RF-SC and FF-SC scenarios, we can also write the GBE of a state, for example, $V_1$ using Eq. \eqref{eqn:GBEnetwork} by including additional transition rates in the RHS of above given GBE equations for $V_1$ under the RF and the FF, respectively, due to departures of a class 1 (1 slice bandwidth) connection on route $r(o=3)$ from two additional network states \{(3,0),(0,3)\} and \{(0,3),(3,0)\}, which are exclusively created because of spectrum conversion at node 2 in Fig. \ref{fig:NetworkStates}(a). 

Under the stationary condition, the network state probabilities  $\boldsymbol{\pi} = [\pi(V_1), \pi(V_2), \cdots, \pi(V_{|\Omega_V|})]$ can be calculated by solving $\boldsymbol{\pi Q}=0$ subject to $\sum_i \pi(V_i)= 1 $, where $\boldsymbol{Q}$ is the transition rate $|\Omega_V| \times |\Omega_V|$ matrix with elements $q_{it}$.  The individual elements $q_{it}, i\neq t$ is obtained by either arrival or departure of a connection $(o,k)$ between each pair of states $V_i$ and $V_t, t\neq i$, which is given by Eq. \eqref{eqn:TRMexact}, and $q_{ii}=-\sum_{t\neq i}q_{it}, i=1,2,\ldots,|\Omega_V|$.  
\begin{equation}
q_{it}=
\begin{cases}
\frac{\lambda_k^o}{|\Gamma_{V_i}^{o,k+}|} & \text{ if } A(i,t,o,k)=k, \text{ for any } (o, k) \\
\mu_k  & \text{ if } A(i,t,o,k)=-k, \text{ for any } (o, k) \\
0 & \text{ otherwise }
\end{cases}
\label{eqn:TRMexact}
\end{equation}
It should be noted that in the FF and FF-SC scenarios, the number of elements in a set $\Gamma_{V_i}^{o,k+}$, i.e., $|\Gamma_{V_i}^{o,k+}|$ is 1 if the allocation function $A(i,t,o,k)=k$ for any pair of $(o,k)$, otherwise it is zero 0.
We can use an LSQR method \cite{paige1982lsqr} or by a successive over-relaxation \cite{young2014iterative} method  to solve  $\boldsymbol{\pi Q}=0$ and $\sum_i \pi(V_i)= 1$, thus the steady state network state distribution $\pi(V_i), i=1,2,\ldots,|\Omega_V|$ can be obtained. 

\par Finally, the overall exact blocking probability in an EON with or without SC is given by ensamble averaging over blocking probability ($BP^o_k$) of all classes $k, k=1,\ldots,K$ on all OD pair requests $o \in \mathcal{O}$ using the blocking identification function $B(i,o,k)$ as follows.
\begin{align} \label{eqn:ExactBPinEON}
BP   &= \frac{\sum_k\sum_o \lambda^o_k BP^o_k}{\sum_k\sum_o\lambda^o_k}  \nonumber \\ 
       &=\frac{\sum_k\sum_o \lambda^o_k \times \left[\sum_{i=1}^{|\Omega_V|}\pi(V_i)\times B(i,o,k) \right]}{\sum_k\sum_o \lambda^o_k} 
\end{align}

\begin{figure}[ht!]
 \centering
\includegraphics[width=0.45\textwidth]{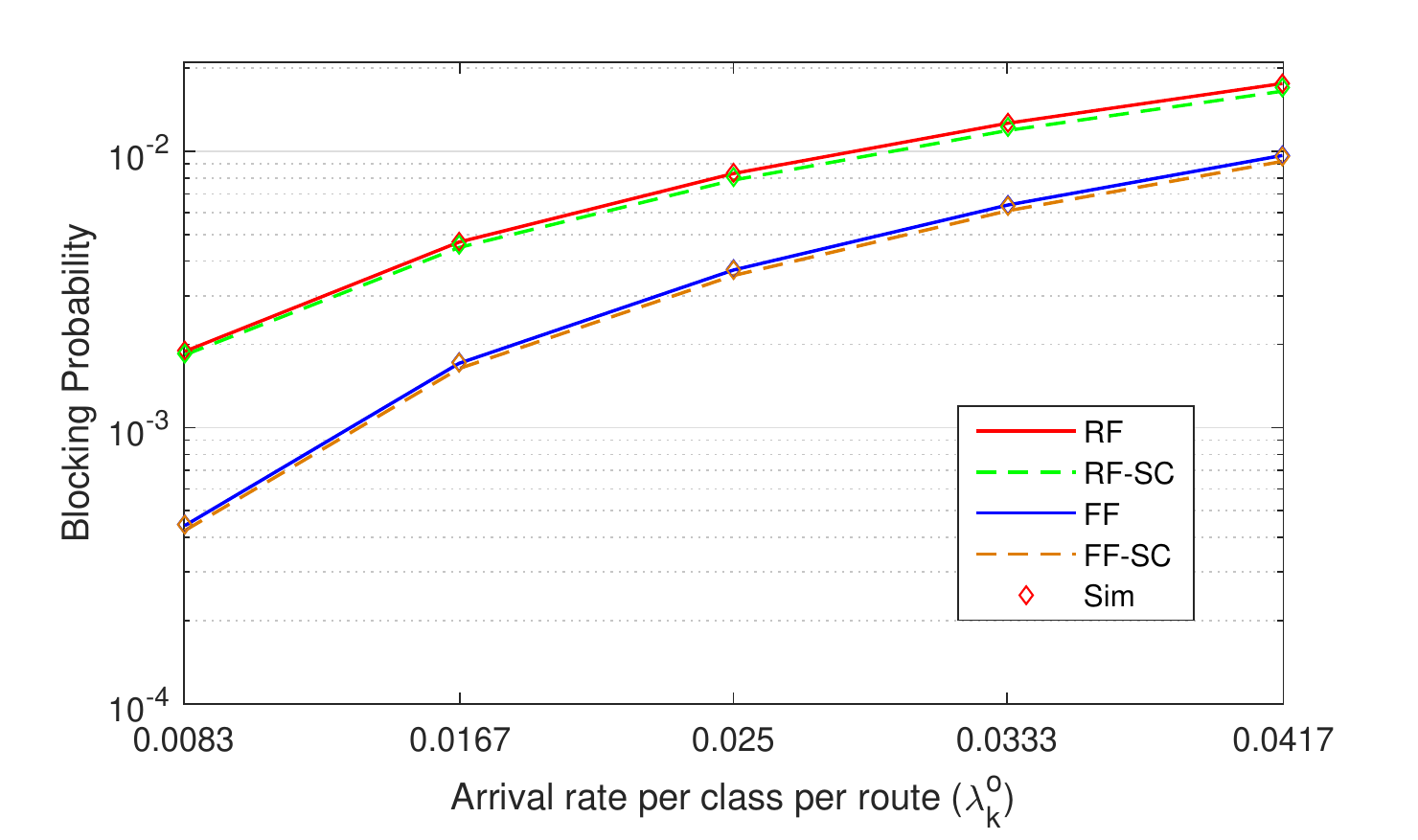}
  \caption{Exact and simulation blocking  results in a 2-hop network with 3 OD pair routes with $C=10$ and demands $d_k=\{3,4\}$ slices, and $\mu_k=1$.}
\label{fig:ExactBP2-hops}
\end{figure}

\begin{figure}[ht!]
 \centering
\includegraphics[width=0.45\textwidth]{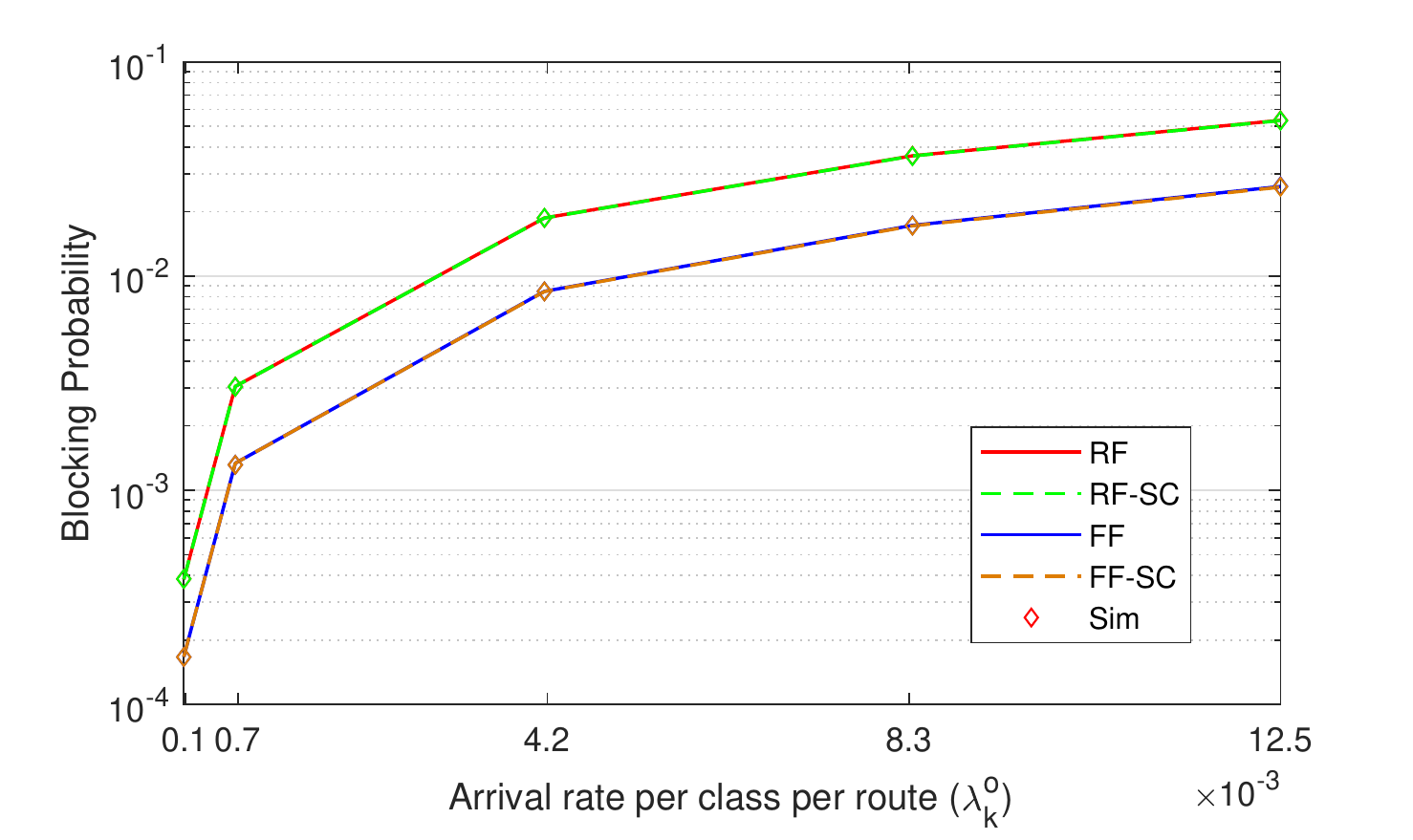}
  \caption{Exact and simulation blocking  results in a 3-node ring network with 6 OD pair routes with $C=7$ and demands $d_k=\{3,4\}$ slices, and $\mu_k=1$.}
\label{fig:ExactBP3-ring}
\end{figure}

We verify the accuracy of our exact model by comparing exact blocking obtained using Eq. \eqref{eqn:ExactBPinEON} under all four operation scenarios to the simulation results in Figs. \ref{fig:ExactBP2-hops} and \ref{fig:ExactBP3-ring} for a 2-link and a 3-node ring topology, respectively. We observe that exact blocking probabilities under all scenarios are very close to the simulation (shown as Sim) results, and the FF exhibits lower blocking than the RF scenario. Furthermore, RF-SC and FF-SC  produce slightly lower blocking than RF and FF, respectively in a small scale 2-link network. However, the blocking reduction due to spectrum conversion would be more visible in large scale links and networks, for which we next propose an approximation model.

\section{Reduced State Model Description} \label{sec:stateDescription}
\par In this Section, we present a reduced link state model in order to tackle the intractability of exact blocking analysis, and identify blocking and non-blocking exact states for a given link occupancy, which will be later used in computing probability of acceptance of a connection request, and also in connection setup rates in the reduced state model. The notations and definitions of some of the parameters used in the model are also listed in Table \ref{table:Notations}.

\begin{table}[t]
\scriptsize
\centering
\caption{Possible spectrum occupancies (states) of a link with capacity $C=7$ slices, demands $d_k=\{3, 4\}$ slices under RF policy.}
\label{table:state description}
\begin{tabular}{@{}c|c|c@{}}
\toprule
Exact link state description                  & Macrostate                                  & Microstate         \\ \midrule
$\textbf{s}=(s_1, s_2, s_3, s_4, s_5, s_6, s_7)$           & \multicolumn{1}{c|}{$\textbf{n}^1=(n_1^1, n_2^1)$}             & $X=x=  \textbf{n}^1\cdot \textbf{d}^T$                \\ \midrule
$\textbf{s}_1=(0, 0, 0, 0, 0, 0, 0)$                          & \multicolumn{1}{c|}{(0,0)}                  & 0                  \\ \midrule
$\textbf{s}_2=(1,\infty,\infty,0,0,0,0)$                & \multicolumn{1}{c|}{\multirow{5}{*}{(1,0)}} & \multirow{5}{*}{3} \\
$\textbf{s}_3=(0,1,\infty,\infty,0,0,0$)                & \multicolumn{1}{c|}{}                       &                    \\
$\textbf{s}_4=(0,0,1,\infty,\infty,0,0$)                & \multicolumn{1}{c|}{}                       &                    \\
$\textbf{s}_5=(0,0,0,1,\infty,\infty,0$)                & \multicolumn{1}{c|}{}                       &                    \\
$\textbf{s}_6=(0,0,0,0,1,\infty,\infty$)                & \multicolumn{1}{c|}{}                       &                    \\ \midrule
$\textbf{s}_7=(1,\infty,\infty,\infty,0,0,0$)           & \multicolumn{1}{c|}{\multirow{4}{*}{(0,1)}} & \multirow{4}{*}{4} \\
$\textbf{s}_8=(0,1,\infty,\infty,\infty,0,0$)           & \multicolumn{1}{c|}{}                       &                    \\
$\textbf{s}_9=(0,0,1,\infty,\infty,\infty,0$)         & \multicolumn{1}{c|}{}                       &                    \\
$\textbf{s}_{10}=(0,0,0,1,\infty,\infty,\infty$)           & \multicolumn{1}{c|}{}                       &                    \\ \midrule
$\textbf{s}_{11}=(1,\infty,\infty,1,\infty,\infty,0$)      & \multicolumn{1}{c|}{\multirow{3}{*}{(2,0)}} & \multirow{3}{*}{6} \\
$\textbf{s}_{12}=(1,\infty,\infty,0,1,\infty,\infty$)      & \multicolumn{1}{c|}{}                       &                    \\
$\textbf{s}_{13}=(0,1,\infty,\infty,1,\infty,\infty$)      & \multicolumn{1}{c|}{}                       &                    \\ \midrule
$\textbf{s}_{14}=(1,\infty,\infty,1,\infty,\infty,\infty$) & \multicolumn{1}{c|}{\multirow{2}{*}{(1,1)}} & \multirow{2}{*}{7} \\
$\textbf{s}_{15}=(1,\infty,\infty,\infty,1,\infty,\infty$) & \multicolumn{1}{c|}{}                       &                    \\ \bottomrule
\end{tabular}
\end{table}

\subsection{Reduced Link State Representation}
The exact network state model is computationally intractable for a medium or large scale links and networks. Thus, it is essential to represent a link state by the number of occupied slices on a link.  Table \ref{table:state description} shows how an exact link state description (formed by a single route) can be equivalently represented by only a few microstates, which represent  the corresponding total occupied slices.  In this 7-slices link example, there are 15 exact link states under the RF policy. However, for example, all 5 exact states having total occupancy of 3 slices ($\textbf{s}_2$ to $\textbf{s}_6$ in first column) are represented by a single Microstate $x=3$, where $X=x= \textbf{n}^1\cdot \textbf{d}^T=\sum_{k=1}^Kn_k^1d_k$, where $\textbf{n}^1 \equiv (n_1^1, \ldots,n_k^1,\ldots, n_K^1)$, and $n_k^1$ is the number of class-$k$ connections of an OD pair request $o=1$. For example, even in a small-scale link with 20 slices and bandwidth demands $d_k = \{3, 4, 5\}$, under the RF policy the number of exact link states is 5885, which could be reduced to 19 with microstates representation. Thus, the reduced state (Microstate) model presents an opportunity to obtain approximate blocking probabilities even for large scale links and networks, since  the maximum number of microstates per link is $C+1$, where $C$ is the number of slices per fiber-link. It should be noted that the term ``state'' is also used in the context of the models (Exact and Microstate), i.e., a state in the Microstate model has the same meaning as a microstate. 

\par Departing from the exact state representation to a microstate representation causes some inaccuracy in finding the connection setup rates in a reduced link state model. The reason is that a microstate  could be represented by different class-dependent blocking and non-blocking exact states formed by one or more routes. For example, a microstate with occupancy $x=3$ slices is represented by five ($\textbf{s}_2$ to $\textbf{s}_6$) different exact link states out of which $\textbf{s}_4$ is a blocking state for both classes of demands requiring 3 and 4 consecutive free slices, and $\textbf{s}_3$ and $\textbf{s}_5$ are additional blocking states for a 4-slice demand. 

Let us first define a set of blocking exact states for an incoming class $k$ request in a microstate $X=x$ by Eq. \eqref{eqn:blocking states}, which can not admit a demand $d_k$ due to the fact that the size of the largest consecutive free slices  ($f_m$)  is not sufficient.
\begin{equation}
\mathbb{B}(x, k) = \{ \textbf{s}_i | d_k > f_m(\textbf{s}_i), \,  \textbf{s}_i \in \Omega_S(x), \, \forall i\}
\label{eqn:blocking states}
\end{equation}
Therefore, a set of non-blocking exact states can be given by 
\begin{equation}
\mathbb{NB}(x,k) = \Omega_S(x) \setminus \mathbb{B}(x,k).
\label{eqn:NB states}
\end{equation}

\par The blocking in a link  happens either due to insufficient free spectrum, referred to as \emph{resource blocking} or due the fragmentation of free spectrum resources, referred to as \emph{fragmentation blocking}.
Fragmentation blocking states ($\mathbb{FB}(x,k)$)  do have enough free slices, but they are scattered and the largest block of consecutive free slices ($f_m(\textbf{s}_i)$) can not satisfy demand $d_k$. Thus, a class-dependent set of fragmentation blocking states corresponding to a microstate $X=x$ is given by 
\begin{equation}
\mathbb{FB}(x,k) = \{ \textbf{s}_i | f_m(\textbf{s}_i) < d_k  \leq C-x, \textbf{s}_i \in \Omega_S(x), \, \forall i\}.
\label{eqn:FB states}
\end{equation}
The set of resource blocking states $\mathbb{RB}(x,k) \!\! \subseteq \! \mathbb{B}(x, k)$ is given by $\mathbb{B}(x, k) \setminus \mathbb{FB}(x, k)$. 

The number of elements in $\Omega_S(x), \mathbb{NB, FB}$, and $\mathbb{RB}$ sets can be obtained using a simple procedure by generating all exact link states under a given spectrum allocation scenario (e.g., RF, RF-SC) using an approach described in Algorithm \ref{algo:ExactNetStates} in Section \ref{sec:ExactModel} for a small scale single-hop (link) network. However, in medium and large scale links with capacity $C>20$, an algorithmic approach would not be useful, since the time and space complexity increase exponentially. Thus, using the inclusion-exclusion principle, we provide analytical expressions for computing the number of exact states ($|\Omega_S(x)|$) on a link with $\mathrm{r}$ traversing routes in Theorem \ref{theorem:Allstates4microstate}, the number of non-blocking exact link states ($|\mathbb{NB}(x,k)|$) in Theorem \ref{theorem:AllNBstates4microstate}, and the number of fragmentation blocking states ($|\mathbb{FB}(x,k)|$) in Theorem \ref{theorem:AllFBstates4microstate}, where the number of free slices $E(x)=C-x$ and the number of connections $N(\textbf{n})=\sum_{k=1}^K\sum_{o=1}^\mathrm{r} n_k^o(\textbf{n})$. The proofs are given in Appendix \ref{sec:Appendix1}.  The number of resource blocking exact link states is $|\mathbb{RB}(x,k)|=|\Omega_S(x)|-|\mathbb{NB}(x,k)|-|\mathbb{FB}(x,k)|$.

\begin{theorem} \label{theorem:Allstates4microstate}
Under the RF policy, the number of exact link states  in a given occupancy state $x$ is 
\begin{equation} \label{eqn:Allstates4microstate}
|\Omega_S(x)| = \sum_{\textbf{n} \in \Omega_S(x)}\frac{N(\textbf{n})!}{\prod_{k=1}^K\prod_{o=1}^\mathrm{r} n_k^o(\textbf{n})!}\times\binom{E(x)+N(\textbf{n})}{N(\textbf{n})}.
\end{equation}
\end{theorem}

\begin{theorem} \label{theorem:AllNBstates4microstate}
Under the RF policy, the number of non-blocking exact link states  for a class $k$ request with demand $d_k$ slices in a given occupancy state $x$ is 
\begin{equation} \label{eqn:AllNBstates4microstate}
|\mathbb{NB}(x,k)|=\sum_{ \textbf{n} \in \Omega_S(x)}W(\textbf{n})\times \frac{N(\textbf{n})!}{\prod_{k=1}^K\prod_{o=1}^\mathrm{r} n_k^o(\textbf{n})!},  \text{ where }
\end{equation}
\begin{equation}
W(\textbf{n}) =  \sum_{i=1}^{N(\textbf{n})+1} (-1)^{i+1} \binom{N(\textbf{n})+1}{i} \binom{E(x)+N(\textbf{n})-id_k}{N(\textbf{n})}. \nonumber
\end{equation}
\end{theorem}

\begin{theorem} \label{theorem:AllFBstates4microstate}
For any policy, the number of fragmentation-blocking exact link states ($|\mathbb{FB}(x,k)|$)  for a class $k$ request with demand $d_k$ slices in a given occupancy state $x$ is
\begin{equation} \label{eqn:AllFBstates4microstate}
|\mathbb{FB}(x,k)| =
\begin{cases}
|\Omega_S(x)|-|\mathbb{NB}(x,k)|, & 0 \leqslant x \leqslant C -d_k \\
0, & \text{otherwise.}
\end{cases}
\end{equation}
\end{theorem}

For example, in Table \ref{table:state description}, the number of exact link states  corresponding to $x=3$, which is represented by a single route and unique $\textbf{n}^1=(1,0)$, is $\frac{1!}{1!}\times \binom{4+1}{1} =5$.  On the other hand, the number of non-blocking exact states for a  class 2 demand ($d_k=4$ slices) in a microstate $x=3$ i.e., $\textbf{n}^1=(1,0)$ and $E(x=3)=4$ slices, is $\binom{1+1}{1}\times\binom{4+1-4}{1}=2$, which can be seen in Table \ref{table:state description}. Notice that the number of exact link states for a microstate as given by Eq. \eqref{eqn:Allstates4microstate} is only valid under the RF spectrum allocation policy, the number of valid exact link states under the FF policy is generally much lower. Furthermore, as described in Section \ref{sec:ExactModel}, exact link states are formed according to a given spectrum allocation policy, classes of demands, and the number of routes that traverses a link under consideration, which further increases the complexity. However, we can reduce the complexity, at the expense of some inaccuracy, by assuming that the exact link states are created only due to a single route. 

To compute approximate blocking in EONs,  all exact model assumptions (in Section \ref{sec:ExactModel}) are considered in the reduced state model, and below we list two additional assumptions:
 \begin{itemize}
\item Spectrum occupancy in a link $j$ is independent from other links $i\neq j; i,j \in \mathcal{J}$, which is called the \emph{independence link} assumption. 
\item All exact link states are formed by a single route, i.e., $\textbf{n} \equiv \textbf{n}^1= (n_1,\ldots, n_K)$, where the route number is omitted.
\end{itemize}

\subsection{Probability of Acceptance of a Connection on a Link} \label{sec:ProbOfAcceptance}
\par Let us now use the above assumptions and definitions of non-blocking and blocking states to reduce an exact link state model into a reduced microstate model. Let $X_j$ be the random variable representing the number of occupied slices ($x_j$) on a link $j$. We define the probability that a link $j$ is in state $x_j$ as
\begin{equation}
\pi_j(x_j) \equiv Pr[X_j=x_j].
\end{equation}
Therefore, using the  link state probability obtained for a given load, the average occupied slices on a link $j$ is  
\begin{equation} \label{eqn:avgOccupiedSlices}
\bar{x}_j= \sum_{0 \leq x_j \leq C} x_j\pi_j(x_j).
\end{equation}
In general, when a route $r$ contains $l$ links, i.e., $r\equiv \{j_1, j_2,\cdots, j_l\}$, we represent the average occupied slices on a route by a vector $\bar{\textbf{x}}_r \equiv (\bar{x}_{j_1}, \bar{x}_{j_2},\cdots, \bar{x}_{j_l})$. Moreover, due to the independence link assumption, the random variables $X_j$'s are independent, i.e., $Pr[X_j=x_j|X_i=x_i] = Pr[X_j=x_j], i \neq j$.

In the reduced model the transition rate from a microstate $X_j=x_j$ to another microstate due to an arrival of a class $k$ request (i.e., connection setup rate) depends on the connection arrival rate and the probability of its acceptance. Noting that only non-blocking exact states corresponding to the microstate $X_j=x_j$ will accept the incoming request, in a single link system (route $r=\{j\}$) the probability of acceptance of a class $k$ connection request with bandwidth $d_k$ in a given occupancy (microstate) $X=x$ (omitting the subscript $j$), i.e., $p_k(x)$ is obtained by  
\begin{multline}
p_k(x) = Pr[Z_r\geq d_k|X=x]  \\
                  =\sum_{\textbf{s}_i\in \Omega_S(x)}\!\!\!\!\!Pr[f_m(\textbf{s}_i)\geq d_k| \textbf{s}_i,X=x]\times Pr[\textbf{s}_i|X=x]
\label{eqn:ProbOfAcceptance}
\end{multline}
where the event \{$Z_r\geq d_k$\} represents  that the route $r$ (here a link $j$) must have equal or more than $d_k$ consecutive free slices to accept a class $k$ request. In a given microstate $X=x$, only a  subset of  exact states representing a microstate $x$  that have sufficient consecutive free slices would accept the class $k$ request ($\forall \textbf{s}_i \in \Omega_S(x): f_m(\textbf{s}_i)\geq d_k$). The first multiplication term in Eq. \eqref{eqn:ProbOfAcceptance} is a probability function resulting in a value 1 if an exact state $\textbf{s}_i$ is a non-blocking state, 0 otherwise.  The second term is the probability of observing the link in an exact state $\textbf{s}_i$  among the set of exact states representing occupancy of $x$ slices, i.e., $\Omega_S(x)$.  However, the calculation of exact state probabilities (for the second term) in a large link is analytically intractable. We need, therefore,  some kind of approximation to  calculate the class- and state-dependent probability of acceptance and connection setup rates. Assuming that \emph{all exact states corresponding to a given microstate have uniform state probability distribution}, i.e., they are equiprobable. Thus, $Pr[\textbf{s}_i|X=x]=1/|\Omega_S(x)|, \forall \textbf{s}_i \in \Omega_S(x)$. We refer to this approximation as an equiprobable exact states (EES) approach. As only non-blocking exact states ($\mathbb{NB}(x,k)$) would allow a class $k$ connection to be accepted in a microstate $x$, therefore, the first multiplication term in Eq. \eqref{eqn:ProbOfAcceptance} would add up to the total number of exact non-blocking states ($|\mathbb{NB}(x,k)|$) representing an occupancy $x$. Thus, the probability of acceptance in  Eq. \eqref{eqn:ProbOfAcceptance} can be approximated in the EES approach as
\begin{equation}
p^{App.EES}_k(x)=\frac{|\mathbb{NB}(x,k)|}{|\Omega_S(x)|}.
 \label{eqn:ApprxProbOfAccept1}
\end{equation} 

The state-dependent per-class connection setup rate in a link is given as the class $k$ arrival rate ($\lambda_k\equiv \lambda_k^o$) multiplied by the probability of  acceptance  of an incoming demand $d_k$ in a microstate $x$, i.e., $\alpha_k(x)=\lambda_k \times p^{App.EES}_k(x)$.

\begin{figure}[t]
 \centering
\includegraphics[width=0.45\textwidth]{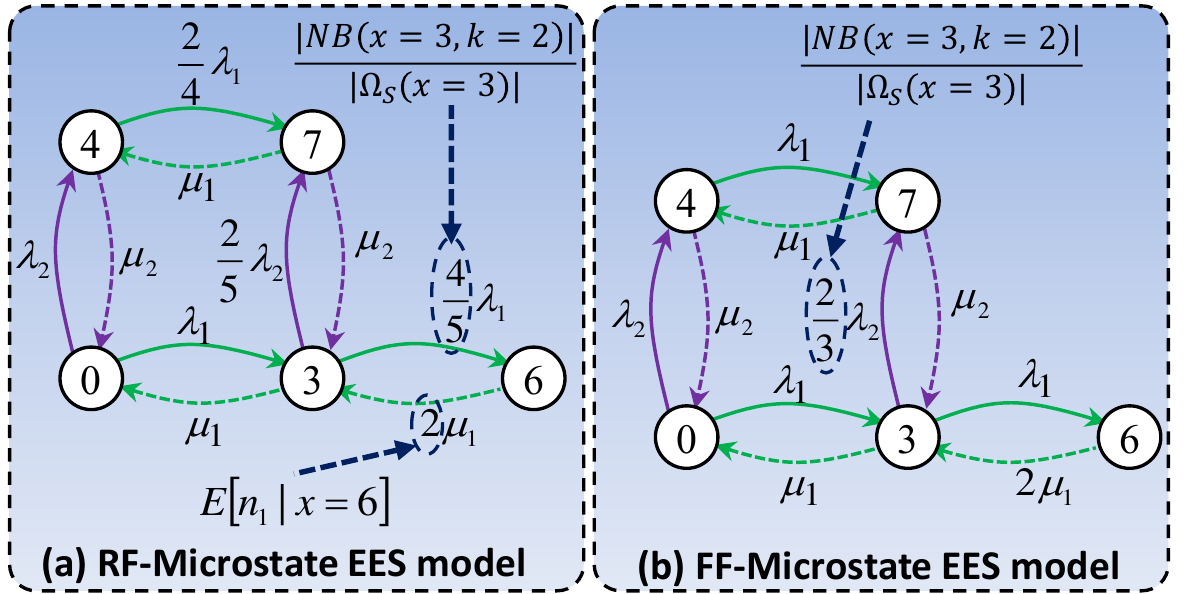}
  \caption{Microstate transition diagram of 7-slice fiber link with two classes of demands $d_k = \{3, 4\}$ slices under RF in (a) and FF in (b) are shown.}
\label{fig:StateTransitions}
\end{figure}

\par To illustrate the transitions and connection setup rates in a reduced state model using the EES approximation, let us consider an example in  Fig. \ref{fig:StateTransitions}, where  the microstate transition diagram of a 7-slice link occupancy is shown with two classes of demands $d_k = \{3, 4\}$ slices under the RF policy in Fig. \ref{fig:StateTransitions}(a), and under the FF policy in Fig. \ref{fig:StateTransitions}(b). As can be seen in Fig. \ref{fig:StateTransitions}(a),  the overall connection setup rate in the empty microstate $x=0$  is $\lambda_1$ and $\lambda_2$ for class-1 (3-slice) and class-2 (4-slice) connection request, respectively, since the corresponding exact empty state  $\textbf{s}_1$  is a non-blocking state for both connection classes, under both RF and FF policies. However, in a microstate $x=3$, which represents 5 different exact states of $\textbf{n}=(1,0)$ in the RF policy, four (two) states are non-blocking for class $k=1 (k=2)$, see Table \ref{table:state description}.  Using the $App.EES$, the connection setup rate for class-1 (class-2) in the microstate $x=3$ is $\frac{4}{5}\lambda_1$ ($\frac{2}{5}\lambda_2$). In contrast, in the FF policy, which allocates only first available slices, generally generates lesser number of exact states as compare to the RF policy. In this example, in the FF policy a microstate $x=3$ is represented by only three exact states $(1,\infty,\infty, 0, 0, 0, 0), (0, 0, 0, 1,\infty,\infty, 0)$ and $(0, 0, 0, 0, 1,\infty,\infty)$, out of which there is only a single exact state  $(0, 0, 0, 1,\infty,\infty, 0)$ that blocks a class-2 demand. Thus, using Eq. \eqref{eqn:ApprxProbOfAccept1} the class-2 connection setup rate in the microstate $x=3$ is $\alpha_2(x=3)=\frac{2}{3}\lambda_2$ as shown in Fig \ref{fig:StateTransitions}(b). Notice that the transition rate from a state occupancy $x=6$ to $x=3$ is $2\mu_1$, since the transition occurs due to the departure of a class-1 connection (3 slices bandwidth), and the expected number of class-1 connections in $x=6$ is 2 in both policies, because $x=6$ is represented by only one $\textbf{n}= (2, 0)$ as shown in Table \ref{table:state description}.

\begin{figure}[t]
 \centering
\includegraphics[width=0.35\textwidth]{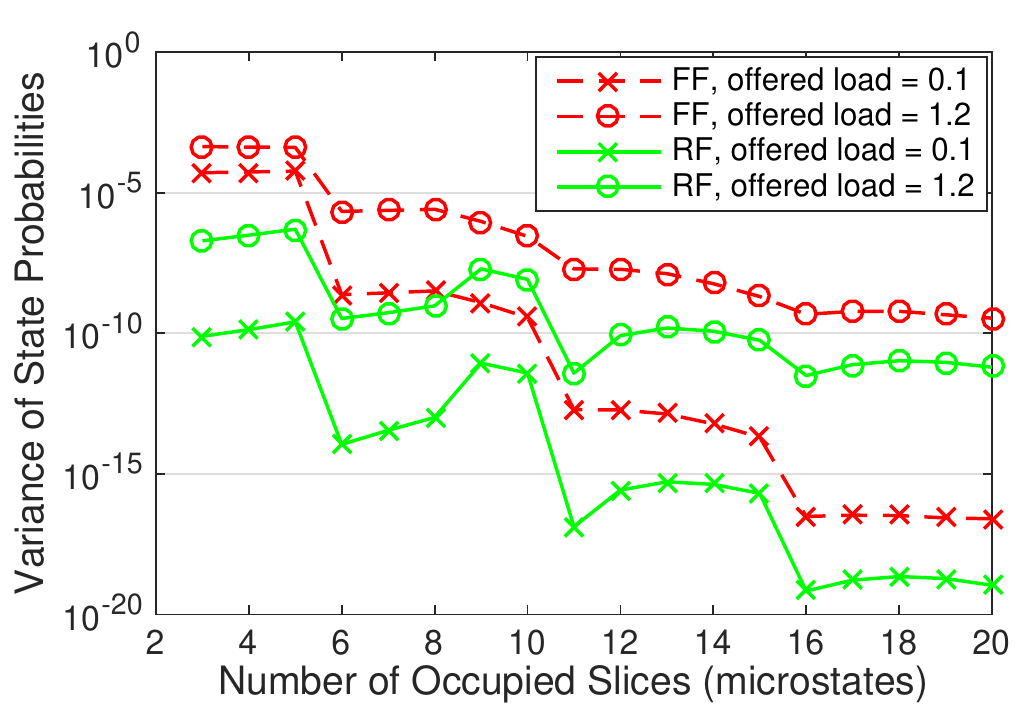}
  \caption{Variance of steady state probabilities of a set of exact states  representing a link occupancy (microstate) for $C=20, d_k=\{3,4,5\}$.}
\label{fig:varianceSP}
\end{figure}

To test the EES assumption corresponding to each microstate (occupied slices)  under both spectrum allocation policies,  we plot the variance of exact state probabilities using the RF and the FF exact models  against total occupied slices in Fig. \ref{fig:varianceSP}. We can see that the variance is non-zero for all microstates, which should not have been the case for the reduced state model to be accurate. Nevertheless, the variance under the RF policy is not as high as the variance obtained in the FF policy. Thus, we make the following observations.

\textit{Observation 1}: The probability of observing an occupancy state ($x$) in its exact states is not eqiprobable. In fact, it varies in its blocking and non-blocking exact states depending on the load, or equivalently, on the average occupied slices $\bar{x}$.

\textit{Observation 2}: In a lower occupancy state ($x \ll \bar{x}$), the probability of observing occupancy $x$ in its non-blocking exact states is more likely than observing in its blocking exact states.

Although both observations are valid for the RF and FF policies, they have huge influence on approximate blocking probabilities obtained under the FF policy, as the distribution of slice occupancy is highly correlated in the FF policy. Thus, the accuracy of $App.EES$ could be compromised under the FF policy.  Therefore, next we consider these two observations and propose a load- and state-dependent SOC approximation which could be used to obtain a relatively more accurate approximate blocking under the FF policy. 


Noting the observations 1 and 2, and the fact that a class $k$ request with demand $d_k$ slices could be accepted in microstates $0 \leqslant x \leqslant C-d_k$, which are represented by non-blocking and fragmentation blocking exact states, we assume that for a class $k$ request in a microstate $x$, the non-blocking exact states are equiprobable, so are the fragmentation blocking exact states.  However, the probability of observing a non-blocking state is higher than observing a fragmentation state in a given microstate ($x$) having   lower occupied slices as compare to the average occupied slices ($x<\bar{x}$). Therefore, let us assume that the probability of observing a non-blocking exact state in a microstate $x$ is in the form $a=\frac{1}{|\Omega_S(x)|}[1+u\exp(-n)]$, and for a fragmentation exact state, it is $b=\frac{1}{|\Omega_S(x)|}[1-v\exp(-n)]; 0\leqslant a,b\leqslant 1$. Additionally, in a given microstate $x, 0 \leqslant x \leqslant C-d_k$, $a\times |\mathbb{NB}(x,k)|+b\times |\mathbb{FB}(x,k)|=1$, and $|\Omega_S(x)|=|\mathbb{NB}(x,k)|+|\mathbb{FB}(x,k)|$. Also noting that only non-blocking states would accept an incoming class $k$ request,  a load-dependent ($\bar{x}$) approximate ($App.SOC$) probability of acceptance of a request in a link may be given by Eq. \eqref{eqn:ApprxProbOfAccept2}.  
\begin{equation}
 p^{App.SOC}_k(x;\bar{x})   =\frac{|\mathbb{NB}(x,k)|}{|\Omega_S(x)|} +\frac{|\mathbb{FB}(x,k)|}{|\Omega_S(x)|}\exp(-\frac{\bar{x}}{C}\times \lvert \ln(\frac{x}{\bar{x}}) \rvert) 
 \label{eqn:ApprxProbOfAccept2}
\end{equation}

Notice that  $ 0 \leq p^{App.SOC}_k(x ; \bar{x}) \leq 1$, since $|\Omega_S(x)|=|\mathbb{NB}(x,k)|+|\mathbb{FB}(x,k)|$ for $0 \leqslant x \leqslant C-d_k$, and $ p^{App.SOC}_k(x,\bar{x})=0$ for resource blocking states, i.e., $C-d_k < x \leqslant C$, since $|\mathbb{NB}(x,k)|=|\mathbb{FB}(x,k)|=0, \text{and}\; |\Omega_S(x)|=|\mathbb{RB}(x,k)|$. As an example, the probability of acceptance of a class $k$ request with demand $d_k \leqslant C$ in an empty state is $p^{App.SOC}_k(0,\bar{x})=1$, since $|\Omega_S(0)| =|\mathbb{NB}(0,k)|=1$, and $|\mathbb{FB}(0,k)|=0$.
The difference between the $App.EES$ and $App.SOC$ is the second factor in Eq. \eqref{eqn:ApprxProbOfAccept2}, which increases the probability of acceptance for lower occupancy states $x$, and decreases it for higher states depending on the  average occupied slices ($\bar{x}$) for a given load. At the same time, the computation of $App.SOC$  requires the knowledge of $\bar{x}$, and vice versa, which makes it a coupled equation. Thus, $\bar{x}$ needs to be computed using an iterative procedure described in Sec. \ref{sec:Methodology4BP}.

\subsection{Probability of Acceptance on a Multi-hop Route} \label{sec:PoA-route}
Let $Z_r$ be the random variable (r. v.) representing the size of the largest continuous and contiguous free slices on a route $r=\{j_1, j_2, \cdots, j_l\}$ without SC, where $j_i$  represents a link on the route $r$. In contrast, under the SC operations,  $Z_r$ would represent the minimum of the largest size of contiguous free slices on each constituent link of route $r$, i.e,  $Z_r=\min(Z_{j_1}, \ldots, Z_{j_l})$, where $Z_{j_i}$ is a r. v. representing the size of the largest contiguous free slices on a link $j_i$, and  the $Z_{j_i}$'s are statistically independent due to the independence link assumption. 
The probability of acceptance of a  connection request with demand $d_k$ slices on a route $r$ without SC can be obtain by extending the single-hop approach in Eq. \eqref{eqn:ProbOfAcceptance} to an $l$-hop route $r$ in a given route occupancy vector $\textbf{x}_r=(x_{j_1},\ldots, x_{j_l})$ with a parameter average route occupancy vector $\bar{\textbf{x}}_r = (\bar{x}_{j_1}, \bar{x}_{j_2},\cdots, \bar{x}_{j_l})$ as follows. 
\begin{multline} \label{eqn:ProbOfAcceptInRoute}
p_k(\textbf{x}_r;\bar{\textbf{x}}_r) = Pr(Z_r\geq d_k \mid X_{j_1}=x_{j_1},..., X_{j_l}=x_{j_l};\bar{\textbf{x}}_r)\\
                         =\!\!\!\sum_{V_t\in \Omega_V(\textbf{x}_r)}\!\!\!\!\!Pr[f_m(\cup_{i=1}^l\textbf{s}^{j_i}_t)\geq d_k| V_t, \textbf{x}_r;\bar{\textbf{x}}_r]\times Pr[V_t|\textbf{x}_r;\bar{\textbf{x}}_r]
\end{multline}
Here, the analogy is similar to that of a single-hop, i.e., in place of an exact state $\textbf{s}$, now a set of $l$ exact link states on the route $r$, belonging to a network state  $V_t \in \Omega_V(\textbf{x}_r)$, i.e., $(\textbf{s}^{j_1}_t,\ldots, \textbf{s}^{j_l}_t), \textbf{s}^{j_i}_t\in \Omega_S(x_{j_i})$  determines whether this set of exact link states  has equal or more than $d_k$ free contiguous and continuous slices or not, which is given by the first multiplication term in Eq. \eqref{eqn:ProbOfAcceptInRoute} with probability 1 or 0. Notice that a slice-wise union operation over a set of $l$ exact link states finds the number of aligned free slices, i.e., the continuity constraint, and the function $f_m(\cdot)$ finds the largest free contiguous slices among the aligned free slices. However, even though we assume that network states with route occupancy $\textbf{x}_r$ are equiprobable, i.e., the second probability term is approximated as $1/|\Omega_V(\textbf{x}_r)|$, an analytical expression for computing the number of exact network states that could accept a request with demand $d_k$ ( i.e., first summation term) is not possible. Moreover, the probability of acceptance of a request with demand $d_k$ on an $l$-hop route with SC in route occupancy vector $\textbf{x}_r$ and average occupancy vector $\bar{\textbf{x}}_r = (\bar{x}_{j_1}, \bar{x}_{j_2},\cdots, \bar{x}_{j_l})$ is given by Eq. \eqref{eqn:ApprxProbOfAcceptInRoute-SC} 
\begin{align} \label{eqn:ApprxProbOfAcceptInRoute-SC}
p_{k,sc}(\textbf{x}_r; \bar{\textbf{x}}_r) =\prod_{i=1}^l Pr(Z_{j_i}\geq d_k|X_{j_i}=x_{j_i}; \bar{x}_{j_i})        
\end{align}
which uses the definition of the r. v. $Z_r$ under the SC operations, i.e., $Z_r=\min(Z_{j_i},\cdots,Z_{j_l})$, and $Z_{j_i}$'s are independent, so  $Pr(\min(Z_{j_1},\cdots, Z_{j_l})\geq d_k)=\prod_{i=1}^l Pr(Z_{j_i}\geq d_k)$.

To address the above issue, we can assume that \emph{the total link occupancy is uniformly distributed over spectrum slices}, which we refer to as $\textit{Uniform}$ approximation. In other words, this approach assumes that spectrum patterns (exact states) are created only by a single-slice demand, thus it ignores a given classes of demands while computing the probability of equal or more than $d_k$  free slices on a route $r$ in both with and without SC operations. For the $\textit{Uniform}$ approach, we derive an analytical expression for computing the probability of acceptance term in Eq. \eqref{eqn:p_k_x_uniform} in Appendix \ref{sec:Appendix2}. Although the $\textit{Uniform}$ approximation considers required RSA constraints in with and without SC operations, it ignores the valid spectrum patterns and fragmentation created due to bandwidth demands, which results in under-estimating the computation of probability of acceptance term. Thus, achieving scalability and accuracy (using valid spectrum patters and RSA constraints) at the same time is hard. Nevertheless, utilizing the independence link assumption and noting that a product-form approximation is also a valid probability distribution \cite{chow1968approximating,peng2013theoretical}, the approximate probability of acceptance of a request on a route with $l$ hops in an EON without SC may be given by a product of all individual probability of acceptance in constituent links on each hop of an $l$-hop route $r$ as below.  
\begin{align} \label{eqn:ApprxProbOfAcceptInRoute}
p^{App.}_k(\textbf{x}_r; \bar{\textbf{x}}_r) &=\left[\prod_{i=1}^l Pr(Z_{j_i}\geq d_k|X_{j_i}=x_{j_i}; \bar{x}_{j_i})\right]^l         
\end{align} 
Notice that the individual link acceptance probability term $Pr(Z_{j_i}\geq d_k|X_{j_i}=x_{j_i}; \bar{x}_{j_i}$) finds the probability that the route $r$ has equal or more than $d_k$ free consecutive slices on each link $j_i \in r$ in corresponding occupancy state $x_{j_i}$. Although Eq. \eqref{eqn:ApprxProbOfAcceptInRoute} does not necessarily ensure that the contiguous free slices are aligned over the route $r$, i.e., the continuity constraint, its effect is partially taken into account by considering only a fraction of all possibilities (using power $l$) of having equal or more than $d_k$ consecutive slices on each link of a route $r$. 

Now, using $App.EES$ in Eq. \eqref{eqn:ApprxProbOfAccept1} and  Eq.  \eqref{eqn:ApprxProbOfAcceptInRoute} we obtain an approximate probability of acceptance of a request on a route $r$ with $l$ hops (without SC) as follows.
 \begin{equation}
p^{App.EES}_k(\textbf{x}_r) = \left[\prod_{i=1}^l \frac{|\mathbb{NB}(x_{j_i},k)|}{|\Omega_S(x_{j_i})|}\right]^l.
 \label{eqn:ApprxProbOfAcceptInRoute1}
\end{equation}
Similarly, for the SC operation modes, using $App.EES$ in Eq.  \eqref{eqn:ApprxProbOfAccept1} and Eq. \eqref{eqn:ApprxProbOfAcceptInRoute-SC}, the probability of acceptance is given on a route $r$ with $l$ hops as follows.
\begin{equation}
p^{App.EES}_{k,sc}(\textbf{x}_r) = \prod_{i=1}^l \frac{|\mathbb{NB}(x_{j_i},k)|}{|\Omega_S(x_{j_i})|}.
 \label{eqn:ApprxProbOfAcceptInRoute1-SC}
\end{equation}

The load-dependent approximate probability of acceptance ($App.SOC$) of a class $k$ request on an $l$-hop route $r$ without SC is obtained by Eq. \eqref{eqn:ApprxProbOfAcceptInRoute2} (using  Eqs. \eqref{eqn:ApprxProbOfAccept2} and  \eqref{eqn:ApprxProbOfAcceptInRoute}). Similarly, under the SC operation, it is given by Eq. \eqref{eqn:ApprxProbOfAcceptInRoute2-SC} using  Eqs. \eqref{eqn:ApprxProbOfAccept2} and  \eqref{eqn:ApprxProbOfAcceptInRoute-SC}.
\begin{equation} \label{eqn:ApprxProbOfAcceptInRoute2}
p^{App.SOC}_k(\textbf{x}_r; \bar{\textbf{x}}_r) = \left[\prod_{i=1}^l p^{App.SOC}_k(x_{j_i}; \bar{x}_{j_i})\right]^l
\end{equation}
\begin{equation} \label{eqn:ApprxProbOfAcceptInRoute2-SC}
p^{App.SOC}_{k,sc}(\textbf{x}_r; \bar{\textbf{x}}_r) = \prod_{i=1}^l p^{App.SOC}_k(x_{j_i}; \bar{x}_{j_i})
\end{equation}

\section{Computing Approximate Blocking Probabilities} \label{sec:Methodology4BP}
In this Section, we present the methodology, as adopted for EONs from the known models for circuit-switched optical networks \cite{birman1996computing,kuppuswamy2009analytic}, to compute approximate blocking probabilities in  EONs with or without spectrum conversion. 

\subsection{Calculating Connection Setup and Departure Rates} \label{subsec:stateprobAndRates}
\par Generally, the class $k$ connection setup rate in a given link state is a function of the given link state occupancy, demand class, and spectrum allocation policy \cite{chung1993computing,kuppuswamy2009analytic}. However, links carry different traffic in EONs, thus taking the average occupied slices ($\bar{x}_j$)  into consideration, and assuming that the time until the next connection is setup on a link $j$ with  $x_j$ occupied slices is exponentially distributed with parameter $\alpha_k^j(x_j)$, the connection setup rate is given by
\begin{equation}
\alpha_k^j(x_j) = \sum_{o:  j \in r(o)} \lambda_k^o Pr(Z_r\geq d_k|X_j=x_j; \bar{\textbf{x}}_r)
\label{eqn:connection setup rate}
\end{equation}
where the summation takes into account the effective arrival rates  of all OD pairs $o \in \mathcal{O}$ whose routes $r(o)$ pass through the link $j$. It should be noted that the effective (reduced) load contribution of an OD pair $o$ on the link $j$ is considered by a probability function $Pr(Z_r\geq d_k|X_j=x_j;\bar{\textbf{x}}_r)$, which depends on the availability of at least required ($d_k$)  free slices (that fulfills the operation-based RSA constraints) on its route $r(o), j \in r(o)$ in a given state with $x_j$ occupied slices on link $j$, and the average occupied slices on its route, i.e., $\bar{\textbf{x}}_r$.  
Let us consider a 2-hop route $r=\{j_1=j, j_2\}$. Then, the probability term is given as
\begin{align}
 & Pr[Z_r\geq d_k|X_j=x_j; \bar{\textbf{x}}_r=(\bar{x}_{j},\bar{x}_{j_2})] \nonumber \\
 &= \sum_{x_{j_2}=0}^C Pr[Z_r\geq d_k|X_j=x_j,X_{j_2}=x_{j_2}; \bar{\textbf{x}}_r] \nonumber \\
 & \,\,\,\,\,\,\,\,   \times Pr[X_{j_2}=x_{j_2}|X_j=x_j] \nonumber \\
 &= \sum_{x_{j_2}=0}^{C-d_k}\pi_{j_2}(x_{j_2}) \times Pr[Z_r \geq d_k|X_j=x_j,X_{j_2}=x_{j_2}; \bar{\textbf{x}}_r]. 
\end{align}

Note that the random variables $X_{j}$ and $ X_{j_2}$ are  independent so $Pr[X_{j_2}=x_{j_2}|X_j=x_j] = Pr[X_{j_2}=x_{j_2}]=\pi_{j_2}(x_{j_2})$.
In general, the above term can be calculated as in \cite{birman1996computing,kuppuswamy2009analytic}, for an OD pair traversing  route $r=\{j_1=j, j_2, j_3,\ldots, j_l\}$ with $l$ hops,  using Eq. \eqref{eqn:ProbOfAcceptEON}.
\begin{multline}\label{eqn:ProbOfAcceptEON}
Pr(Z_r\geq d_k|X_j=x_j;\bar{\textbf{x}}_r)=\!\!\!\sum_{x_{j_2}=0}^{C-d_k}\!\!\!\cdots\!\!\!\sum_{x_{j_l}=0}^{C-d_k}\!\!\!\pi_{j_2}(x_{j_2})\!\cdots\pi_{j_l}(x_{j_l})\\ 
\times Pr(Z_r\geq d_k|X_j\!=\!x_j, X_{j_2}\!=\!x_{j_2},\cdots, X_{j_l}\!=\!x_{j_l}; \bar{\textbf{x}}_r)
\end{multline}
In Eq. \eqref{eqn:ProbOfAcceptEON},  the term after multiplication is referred as  the probability of acceptance of a connection path request with demand $d_k$,  i.e., $p_k(\textbf{x}_r;\bar{\textbf{x}}_r)$, and it can be approximately given under various scenarios with and without SC by the Uniform, EES, and SOC approaches, as shown in Section \ref{sec:PoA-route}.   


The expected departure rate of a class $k$ connection in a state $x_j$ is obtained by Eq. \eqref{eqn:expected departure rate}
\begin{align}
\gamma_k^j(x_j)  &=  \mu_k\times E[n_k|X_j=x_j] \nonumber \\
                           &= \mu_k\times \frac{1}{|\textbf{n}(x_j)|}\sum_{\textbf{n}:\textbf{d}\cdot\textbf{n}^T=x_j}n_k(\textbf{n})
\label{eqn:expected departure rate}
\end{align}
where $E[n_k|X_j=x_j]$ is the expected number of class $k$ connections in the state $x_j$, which is given by assuming all $\textbf{n}$ that results into the same $x_j$ (i.e., $\textbf{n}(x_j)$) have uniform distribution  $\frac{1}{|\textbf{n}(x)|}$, and $n_k(\textbf{n})$ is the number of class $k$ connections in $\textbf{n}$ (remember that in the reduced state model, we assumed that $\textbf{n}\equiv \textbf{n}^1=(n_1, \ldots, n_k)$). 

\subsection{Computing Blocking in EONs} \label{sec:BPw/oDF}
Before we calculate blocking probability in EONs, we need to find out the steady state link occupancy distribution $\pi(x_j), 0\leq x_j \leq C$ for all links $j \in \mathcal{J}$, which can be obtained by solving a set of global balance equations (GBEs) with a normalizing condition $\sum_{x_j=0}^C\pi(x_j)=1$ for each link $j \in \mathcal{J}$. The GBE of a microstate $X_j=x_j$ is  given as 
\begin{multline} \label{eqn:GBEnormalApprox}
\!\!\sum_{k=1}^K  \left( \alpha_k(x_j)+\gamma_k(x_j) \right)\pi(x_j)=\!\!\!\!\! \sum_{k=1, d_k \leq x_j\leq C}^K \!\!\!\!\!\!\!\!\!\!\!\!\! \alpha_k(x_j-d_k) \pi(x_j-d_k) \\
+  \sum_{k=1, 0 \leq x_j\leq C-d_k}^K \!\!\!\!\!\! \gamma_k(x_j+d_k) \pi(x_j+d_k)
\end{multline}
where, LHS represents the output flow rate from a microstate $X_j=x_j$ taking into account the  connection setup rate $\alpha_k(x_j)$ and departure of connection(s) with expected rate $\gamma_k (x_j)$, while the RHS represents input flow rate into the microstate $x_j$ from other state(s) $x_j-d_k$ ($x_j+d_k$) due to an arrival (departure) of a class $k$ connection demand of $d_k$ slices. Thus, for example, using Eq. \eqref{eqn:GBEnormalApprox} the GBE of a microstate $x=3$ in a link in Fig. \ref{fig:StateTransitions}(a) can be written as $(\frac{4}{5}\lambda_1+\frac{2}{5}\lambda_2 + \mu_1)\pi(x=3)=\lambda_1 \pi(x=0)+ 2\mu_1 \pi(x=6) + \mu_2 \pi(x=7)$. Here, the LHS of the GBE of the state $x=3$ takes into the account of a 3-slice (4-slice) demand arrival in $x=3$ with effective connection setup rate $4\lambda_1/5$ ($2\lambda_2/5$), and the RHS terms are due to an arrival in $x=0$, and departures in states $x=6$ and $x=7$. 
 The above linear equations \eqref{eqn:GBEnormalApprox} for all microstates $0 \leq x_j \leq C$ and links $j \in \mathcal{J}$ can also be solved by the LSQR method \cite{paige1982lsqr} to obtain the steady state link occupancy distribution $\pi(x_j)$. Remember that  the connection setup rate ($\alpha_k(x_j)$) depends on  state probabilities ($\pi(x_j)$), thus an iterative procedure is required to obtain steady state link occupancy probabilities. 

\par The class $k$ blocking probability in an EON with or without SC is given by ensamble averaging over class $k$ blocking probability of all OD pair requests $o \in \mathcal{O}$ opting for respective route $r(o)$, and using Eq. \eqref{eqn:ProbOfAcceptEON}, blocking probability of class $k$ bandwidth requests on an OD pair $o$ with bandwidth $d_k$  in an EON can be given as follows.
\begin{align} \label{eqn:BPinEON}
BP_k^o  &= Pr[Z_r<d_k] = 1- Pr[Z_r \geq d_k]  \nonumber \\
&=1- \sum_{x_j=0}^{C-d_k} Pr[Z_r\geq d_k|X_j=x_j; \bar{\textbf{x}}_r] \times Pr[X_j=x_j]  \nonumber \\
&= 1- \sum_{x_j=0}^{C-d_k}\sum_{x_{j_2}=0}^{C-d_k}\cdots \sum_{x_{j_l}=0}^{C-d_k}(\pi_j(x_j)\pi_{j_2}(x_{j_2})\cdots \pi_{j_l}(x_{j_l}) \nonumber \\
 & \,\,\,\,\,\,\,\,\,\,\,\,\,\,\,\,\, \times p_k (\textbf{x}_r; \bar{\textbf{x}}_r))
\end{align}

Notice that blocking probability in a single-hop (link) system is also given by Eq. \eqref{eqn:BPinEON} by omitting $j_i, i\neq 1$ summation  and related state probabilities terms, and  $p_k (\textbf{x}_r; \bar{\textbf{x}}_r)$ is simplified to $ p_k (x_j; \bar{x}_j)$, by setting the number of hops $l=1$ in Eqs. \eqref{eqn:ApprxProbOfAcceptInRoute1}--\eqref{eqn:ApprxProbOfAcceptInRoute2-SC}. More importantly, we use Eq. \eqref{eqn:BPinEON} to obtain approximate blocking probabilities in an EON with and without SC, by calculating $p_k (\textbf{x}_r; \bar{\textbf{x}}_r)$ separately in with and without SC operation modes.

\subsection{Algorithm for Computing Blocking Probabilities in EONs} \label{subsec:IterativeAlgo}
\par The calculation of approximate blocking probability per class per OD pair ($BP_k^o$)  requires the information of steady state link occupancy probabilities ($\pi_j(x_j)$) of each traversed link of a route $r(o)$; and these probabilities $\pi_j(x_j)$   can be obtained by solving the nonlinear coupled equations in Eq. \eqref{eqn:GBEnormalApprox}. However, these nonlinear coupled equations, which are a function of $\boldsymbol{\alpha}$ and $\boldsymbol{\pi}$, could be made linear by repeated substitution or iterative procedure as follows \cite{kuppuswamy2009analytic}.

\begin{enumerate}[1)]
\item For all classes $k \in \{1, 2, \cdots, K\}$ and  OD pairs $o \in \mathcal{O}$, initialize blocking probabilities $\hat{BP}_k^o=0$, and set  $\alpha_k^j(\cdot)$ for each link $j \in \mathcal{J}$ as $\sum_{o:  j \in r(o)} \lambda_k^o$, and $\bar{x}_j=C/2$.
\item Determine the link state occupancy distribution for valid  $x_j, 0 \leqslant x_j \leqslant C$ as $\boldsymbol{\pi_j}=\left[\pi_j(x_j=0),\ldots, \pi_j(x_j=C)\right]$ for each link $j \in \mathcal{J}$ by solving $\boldsymbol{\pi_j}\cdot \boldsymbol{Q_j}=0$ and $\sum_{x_j=0}^C \pi_j(x_j)=1$ using LSQR method \cite{paige1982lsqr}. Here, $\boldsymbol{Q_j}$ is the transition rate matrix formed by the connection setup rates $\alpha_k^j(\cdot)$ and the expected departure rates $\gamma_k^j(\cdot)$. 
\item Calculate $\bar{x}_j$ by Eq. \eqref{eqn:avgOccupiedSlices} and per-class connection setup rate $\alpha_k^j(\cdot) \forall j \in \mathcal{J}, \forall k \in \{1, 2, \cdots, K\}$ using Eq. \eqref{eqn:connection setup rate}. 
\item Calculate $BP_k^o, \forall$ OD pairs $o$ and classes $k$  by Eq. \eqref{eqn:BPinEON}. 
\item If $\max_{o,k}|\hat{BP}_k^o - BP_k^o|<\epsilon$ then terminate. Else, let $\hat{BP}_k^o = BP_k^o$ and go to step (2).
\end{enumerate}

\begin{table}[t]
\centering
\scriptsize
\caption{Comparing various approximation methods.}
\label{my-label}
\begin{tabular}{|c|c|c|c|c|c|p{0.5cm}|p{0.5cm}|}
\hline
\multirow{3}{*}{Scenarios} & \multicolumn{7}{c|}{$C=10, d_k=\{3, 4\}, \lambda_k^o=0.1/6, \mu_k=1$} \\ \cline{2-8} 
         &  \multirow{2}{*}{Exact}             & \multirow{2}{*}{Sim.}              & \multirow{2}{*}{Uniform}   & \multirow{2}{*}{EES}   & \multirow{2}{*}{SOC}   & \multicolumn{2}{c|}{\cite{beyranvand2014analytical}}  \\ \cline{7-8}
             &               &               &    &     &     &  App.1 & App.2 \\ \hline
RF    &   4.7$e$-3      &  4.7$e$-3         & \multirow{2}{*}{2.7$e$-2}        &  6.5$e$-3       &    1.9$e$-3      &  \multirow{4}{*}{3.6$e$-4}  & \multirow{2}{*}{4.5$e$-4} \\ \cline{1-3}  \cline{5-6}
FF    &   1.7$e$-3      &  1.7$e$-3         &         &   8.7$e$-3      &    2.1$e$-3       &    & \\ \cline{1-6} \cline{8-8}
RF-SC   & 4.6$e$-3      &  4.5$e$-3         & \multirow{2}{*}{2.7$e$-2}        &  5.1$e$-3       &    1.7$e$-3      &    &\multirow{2}{*}{2.9$e$-4} \\ \cline{1-3} \cline{5-6}
FF-SC    & 1.7$e$-3      &  1.7$e$-3        &          &  6.7$e$-3       &    1.8$e$-3      &    & \\ \hline
\end{tabular}

\begin{tabular}{|c|c|c|c|c|c|c|c|}
\hline
\multirow{3}{*}{Scenarios} & \multicolumn{6}{c|}{$C=100, d_k=\{3, 4, 6\}, \lambda_k^o=9/9,  \mu_k=1$} \\ \cline{2-7} 
              &\multirow{2}{*}{Sim.}              & \multirow{2}{*}{Uniform}   & \multirow{2}{*}{EES}   & \multirow{2}{*}{SOC}  & \multicolumn{2}{c|}{\cite{beyranvand2014analytical}}  \\ \cline{6-7}
                &               &               &    &        &  App.1 & App.2 \\ \hline
RF          & 4.5$e$-4         & \multirow{2}{*}{3.6$e$-2}         &   \multirow{2}{*}{2.9$e$-4}      &    \multirow{2}{*}{ 9.8$e$-5}      &  \multirow{4}{*}{ 1.7$e$-11}  & \multirow{2}{*}{ 9.0$e$-5} \\ \cline{1-2}   
FF         &   6.3$e$-6        &          &        &           &    &   \\ \cline{1-5} \cline{7-7}
RF-SC   & 1.9$e$-4          & \multirow{2}{*}{1.3$e$-2}       &  \multirow{2}{*}{2.1$e$-4}       &    \multirow{2}{*}{ 5.2$e$-5}      &    &  \multirow{2}{*}{ 3.3$e$-8} \\ \cline{1-2} 
FF-SC    &   4.7$e$-6             &          &         &          &    &  \\ \hline
\end{tabular}
\end{table}

To illustrate the effectiveness of approximation approaches, we compare blocking probabilities (BPs) obtained by various approximations, including two approximations (Kaufman as App.1 and Binomial as App.2)\footnote{In \cite{beyranvand2014analytical}, App.1 BP is obtained by Eq. 16 (page 1627), and for App.2, using a slice occupancy probability $\rho$=$\frac{1}{C}\sum_{j=0}^Cjg(j)$ (Eq. 18), the probability of acceptance terms ($f(C,i)$) in Eqs. 17 and 21 \cite{beyranvand2014analytical}  are correctly given in \cite{peng2013theoretical}.} from \cite{beyranvand2014analytical}, under various operation modes, classes of demands, and link capacity, and validate them by exact and/or simulation results (Sim.) in a 2-link network with 3 OD pair routes (shown in Fig. \ref{fig:NetworkStates}a) over which connection requests arrive according to a Poisson process. Also note that for a small scale scenario ($C=10$), in addition to exact blocking results we provide approximate blocking probability under all four scenarios (RF, FF, RF-SC, and FF-SC) using the EES and SOC approximation approaches by generating exact link states (by assuming a single traversing route) using the Algorithm \ref{algo:ExactNetStates} separately for the RF and FF spectrum allocation policies.  We observe that the $\textit{Uniform}$  approximation yields very high  blocking probabilities in EONs with and without SC. The reason is that it ignores the spectrum fragmentation created by a given classes of demands. On the other hand, the EES approach provides good approximation for the RF policy with and without SC irrespective of the link capacity. Interestingly, the SOC approach is better than the EES approach for the FF policy, thus it can be used to estimate approximate blocking probabity under the FF policy with and without SC. As expected, the approximate (App.1) blocking probability obtained using the Kaufman link state distribution formula in \cite{beyranvand2014analytical} do not match any scenarios, and the Binomial approach (App.2) is a relatively better approach than the Kaufman approach for lower loads and for small scale EONs, as observed in \cite{beyranvand2014analytical}. Thus, to evaluate blocking probability in EONs for all scenarios, we present mainly the EES and the SOC approximations in next section.

\begin{table*}[ht!]
\centering
\caption{Blocking probability in various sized link with different capacities and class of demands.}
\label{table:linkresults}
\begin{tabular}{@{}cccccccccccccc@{}}
\toprule
\multirow{3}{*} & \multicolumn{12}{c}{$C=10, d_k=\{3,4\}$ }                             \\ \cmidrule(l){2-13} 
Scenarios        & \multicolumn{4}{c}{offered load = 0.1} & \multicolumn{4}{c}{offered load = 0.6} & \multicolumn{4}{c}{offered load = 1.2}\\ \cmidrule(l){2-13} 
              & Exact          & Sim.              & App.EES      & App.SOC       & Exact              & Sim.           & App.EES       & App.SOC      & Exact           & Sim.            & App.EES          & App.SOC  \\ \midrule
RF           &   6.8$e$-3   &  6.8$e$-3    &  6.8$e$-3   & 2.7$e$-3   &  9.4$e$-2      &  9.4$e$-2     &  9.5$e$-2 & 6.7$e$-2 &  2.2$e$-1      &  2.2$e$-1     &   2.2$e$-1   & 1.7$e$-1     \\ [0.1cm] 
FF           &    2.9$e$-3   & 2.9$e$-3    &   8.3$e$-3  &  2.8$e$-3   &   6.9$e$-2     &  6.8$e$-2     &  8.6$e$-2 & 6.4$e$-2 &   1.8$e$-1     &  1.8$e$-1     &  2.0$e$-1    & 1.7$e$-1      \\  \bottomrule
\end{tabular}
\end{table*}

\begin{table*}[ht!]
\vspace{-0.2cm}
\centering
\begin{tabular}{@{}ccccccccccccc@{}}
\toprule
 & \multicolumn{12}{c}{$C=100, d_k=\{3, 4, 6\}$ }  \\ \cmidrule(l){2-13} 
Scenarios  & \multicolumn{3}{c}{offered load = 8} & \multicolumn{3}{c}{offered load = 12} & \multicolumn{3}{c}{offered load = 16} & \multicolumn{3}{c}{offered load = 20} \\ \cmidrule(l){2-13} 
 & Sim. & App.EES & App.SOC & Sim. & App.EES & App.SOC & Sim. & app.1 & app.2 & Sim & App.EES & App.SOC \\ \cmidrule(l){1-13} 
 RF        &  1.6$e$-3 & \multirow{2}{*}{1.8$e$-3} &\multirow{2}{*}{4.9$e$-4}    & 2.3$e$-2 & \multirow{2}{*}{2.5$e$-2}  &\multirow{2}{*}{8.5$e$-3}   & 8.1$e$-2  & \multirow{2}{*}{8.7$e$-2} &\multirow{2}{*}{3.8$e$-2}&  1.6$e$-1   & \multirow{2}{*}{ 1.6$e$-1}    &\multirow{2}{*}{ 9.7$e$-2}  \\ [0.1cm]
 FF        &   1.1$e$-4 &   &   &  7.2$e$-3 &   &   &  4.8$e$-2  &  & &   1.2$e$-1   &     &   \\  \bottomrule
%
 Scenarios     &\multicolumn{12}{c}{$C=200, d_k=\{4, 6, 10\}$ }  \\ \midrule   
 RF         &  8.4$e$-5 & \multirow{2}{*}{5.6$e$-6} &\multirow{2}{*}{1.4$e$-6}    & 3.4$e$-3 & \multirow{2}{*}{6.7$e$-4}  &\multirow{2}{*}{2.4$e$-4}   & 2.3$e$-2  & \multirow{2}{*}{1.0$e$-2} &\multirow{2}{*}{4.4$e$-3} &  6.5$e$-2   & \multirow{2}{*}{ 4.6$e$-2}    &\multirow{2}{*}{ 2.2$e$-2} \\ [0.1cm]
  FF        &   2.0$e$-7 &   &   &  1.3$e$-4 &   &   &  3.7$e$-3  &  &    &  2.4$e$-2     &     &  \\ \bottomrule
\end{tabular}
\end{table*}

\section{Numerical and Simulation Results}\label{sec:evaluation}
\par In this section, we investigate the accuracy of approximate blocking probabilities (obtained by $App.EES$ and $App.SOC$) by comparing them with the discrete event simulation results obtained in a unidirectional fiber link, a 14-node (42 links) NSF network, and a 6-node ring network. Additionally, for a small capacity fiber-link  ($C=10$) we compare $App.EES$ and $App.SOC$ BPs  with the exact blocking results obtained by Eq. \eqref{eqn:ExactBPinEON} under two different spectrum allocation policies,  RF and FF. Furthermore, we compare blocking in a multi-hop EON without spectrum conversion (SC) for the RF and the FF policies (simply shown as RF and FF)  to the blocking obtained in the same network enabled with SC  (shown as RF-SC and FF-SC). For the RF and the RF-SC scenarios, total exact link states, non-blocking, and fragmentation blocking link states are obtained by Eqs. \eqref{eqn:Allstates4microstate}--\eqref{eqn:AllFBstates4microstate}, and they are used in calculating probability of acceptance and BP results under $App.EES$ and $App.SOC$.  On the other hand, for the FF and the FF-SC scenarios, the numbers of these link states are given by generating all valid exact link states (with a single traversing route) obtained by the Algorithm \ref{algo:ExactNetStates}  under the FF policy in small scale links and networks. For medium and large scale links and networks ($C>10$), finding the number of non-blocking, blocking and total exact states under the FF policy are computationally challenging, therefore, the simulation results obtained for the FF and the FF-SC scenarios are compared with approximate BPs obtained under the RF and the RF-SC, respectively. 

\par Blocking results are depicted versus \emph{offered load}, which is defined as $\sum_k \sum_o \frac{\lambda_k^o}{\mu_k}$, where $o\in\mathcal{O}, k=1,2,\cdots,K$. We assume that the service (holding) times of connection requests between an OD pair are exponentially distributed with mean $1/\mu_k=1$ unit \cite{yu2013first,beyranvand2014analytical}, and per-class, per OD pair connection requests arrive according to a Poisson process with uniformly distributed rate  $\lambda_k^o$= \emph{offered load}$/(|\mathcal{O}|\times K)$. We compute the average BP as $\sum_k\sum_o\lambda_k^oBP^o_k/\sum_k\sum_o\lambda_k^o$, i.e., by ensemble averaging over BP of all OD pairs $o \in \mathcal{O}$ and classes $k= 1, 2, \cdots, K$. All exact and simulation results presented here consider both spectrum contiguity and spectrum continuity constraints for the RF and FF scenarios (i.e., without SC), and only contiguity constraint is considered for the RF-SC and FF-SC scenarios.  We generated $10^7$ connection requests to simulate small and large scale link as well as EONs. We consider different bandwidth demands $d_k=\{3, 4, 6, 10\}$ slices in EONs, which are equivalent to lightpaths with a guardband on both sides and supporting different bit rates:10, 40, 100 and 400 gigabit per second (Gb/s)  using different modulation formats $M=2$ (e.g., QPSK) and $M=4$ (e.g., DP-QPSK), and slice width granularity is 12.5 GHz \cite{singh2017analytical,singh2017efficient}.

\subsection{Link Model}
\par Note that for a unidirectional link only one OD pair  exists, thus  $\lambda_k= \lambda_k^o=$ \emph{offered load}$/K$. Table \ref{table:linkresults} presents exact (denoted by Exact), verifying simulation (Sim.) and approximate ($App.EES$ and $App.SOC$) BPs in a link with different link capacities and set of demands for various offered loads. In a small scale scenario, $C=10, d_k=\{3, 4\}$ slices, it can be seen that verifying simulation results are very close to the exact results, thus in a large scale link or EONs where the exact solution is intractable, simulation results can be used to verify the approximate solutions.  From Table \ref{table:linkresults} we observe that approximate BP results obtained by $App.EES$ are also very close to the exact solutions under both RF and FF spectrum allocation policies. Interestingly, unlike the RF policy, BPs obtained by $App.SOC$ in the FF policy  is even closer to the exact BPs than that of $App.EES$. This is due the fact that the variance of exact state probabilities is much higher in the FF policy (see Fig. \ref{fig:varianceSP}), and unlike $App.EES$, $App.SOC$ tries to consider spectrum occupancy correlation by assigning higher acceptance probability for non-blocking states for lower occupancy states ($x < \bar{x}$) using the average occupied slices parameter ($\bar{x}$). Furthermore, as expected, BP under the FF policy is lower than that of RF due to the lower spectrum fragmentation \cite{rosa2015statistical,singh2016defragmentation}, thus it is suitable in large scale links or networks where the goal is to increase the number of served connections. However, computing approximate yet accurate BPs under the FF policy is not easy. Nevertheless, in the medium and large scale links ($C=100$ and $C=200$), we obtain BPs in $App.EES$ and $App.SOC$ by computing probability of acceptance in Eqs. \eqref{eqn:ApprxProbOfAccept1} and \eqref{eqn:ApprxProbOfAccept2}, respectively using Eqs.  \eqref{eqn:Allstates4microstate}--\eqref{eqn:AllFBstates4microstate}, and show the simulations obtained under the RF policy and the FF policy separately. We see that BPs given by $App.EES$ can be treated as  approximate BPs for the RF policy, and $App.SOC$ for the FF policy, since BPs obtained by $App.SOC$ are very close to BPs given by simulation results under the FF policy. Additionally, we observe that the approximate BPs deviate for a large scale link ($C=200$) under lower offered loads. The reason is that approximate BPs involve numerical computation, e.g., $\binom n k$, and for a larger $n$ (w.r.t. $k$), the computation is slightly error prone even in a powerful mathematical software (e.g., Matlab), and also due to the fact that at lower loads all possible link states in a large capacity link could not be sufficiently visited even simulating with large number of events, so simulation results might also not be very accurate.

\par In Table \ref{table:runTime} we present computational run time of the exact and approximate solutions, which are obtained on a PC with Intel 6-core i7 3.20 GHz processor with 32 GB RAM. We observe that the run times for the exact and approximate solutions are nearly same for both RF and FF spectrum allocation policies  in a link with smaller capacity ($C=10$). However, when capacity of the link increases to $C=20$, finding exact solutions becomes extremely time and resource (memory) consuming and increases exponentially with respect to capacity $C$. On the other hand, the $App.EES$ and $App.SOC$ solutions under the RF policy takes only a fraction of seconds for $C=20$, and a few seconds for $C=100$ and $C=200$. Interestingly, $App.EES$ and $App.SOC$ solutions for the FF policy is also possible to obtain by generating all possible exact states using the Algorithm \ref{algo:ExactNetStates} for capacity $C=20$, but it is time consuming.  Also note that BP obtained by $App.EES$ and $App.SOC$ for multi-hop EONs with and without SC depends on various factors, including number of OD pair routes, number of links, link capacity, and traffic classes.
\begin{table}[t]
\centering
\caption{The computational run time of different solutions (in seconds) }
\label{table:runTime}
\begin{tabular}{@{}lcccccc@{}}
\toprule
&\multicolumn{3}{c}{$C=10$} & \multicolumn{3}{c}{$C=20$}                               \\ 
&\multicolumn{3}{c}{$d_k=\{3,4\}$} & \multicolumn{3}{c}{$d_k=\{3,4,6\}$}     \\ \cmidrule(l){2-7}
       & Exact  & App.EES & App.SOC & Exact  & App.EES & App.SOC \\ \midrule
RF     &   0.123     &    0.131  & 0.163  &       13971.6                  &        0.369   &     1.021                       \\
FF     &   0.125    &  0.134      &  0.173  &     1746.7                  &        1717.3             &       1719.5         \\ [0.1cm] \bottomrule
\end{tabular}
\end{table}
\begin{table}[t]
\vspace{-0.3cm}
\centering
\begin{tabular}{@{}lcccccccc@{}}
\toprule
 & App.EES (RF) & App.SOC (RF)  \\
\midrule
$C=100, d_k=\{3,4,6\}$ & 6.432 & 7.058    \\[0.2cm]
$C=200, d_k=\{4,6,10\}$ & 40.40 & 40.92   \\ \bottomrule
\end{tabular}
\end{table}

\begin{table*}[ht!]
\centering
\caption{The BP in a NSFnet under various capacities and class of demands.} 
\label{table:NSFresults}
\begin{tabular}{@{}ccccccccccccc@{}}
\toprule
\multirow{3}{*}{} & \multicolumn{12}{c}{$C=10, d_k=\{3,4\}$ }                             \\ \cmidrule(l){2-13} 

Scenarios        & \multicolumn{3}{c}{offered load=0.1} & \multicolumn{3}{c}{offered load=0.6} & \multicolumn{3}{c}{offered load=1.2} & \multicolumn{3}{c}{offered load=7.2}\\ \cmidrule(l){2-13} 
               & Sim. & App.EES & App.SOC & Sim. & App.EES & App.SOC & Sim. & App.EES & App.SOC & Sim & App.EES & App.SOC \\ \cmidrule(l){1-13}                                          
RF     &    3.9$e$-4           &    9.8$e$-4    &  3.5$e$-5         &     3.3$e$-3           &     6.6$e$-3 &  1.1$e$-3            &  8.4$e$-3     &    1.5$e$-2  &  4.0$e$-3           &   1.1$e$-1            &   1.3$e$-1 &  8.6$e$-2 \\ [0.1cm]  
FF      &    2.6$e$-5           &   1.5$e$-3      &  4.6$e$-5      &    9.6$e$-4           &  9.4$e$-3  &   1.3$e$-3          &  3.7$e$-3             &  1.9$e$-2   &   4.5$e$-3       &   8.1$e$-2            &  1.2$e$-1  &  8.1$e$-2\\  [0.1cm]
RF-SC      &  3.6$e$-4       &   4.8$e$-4     &    2.4$e$-5       &     3.0$e$-3           &    4.1$e$-3  &    8.0$e$-4          &   7.4$e$-3             &    1.1$e$-2 &       3.0$e$-3 &    9.2$e$-2            & 1.3$e$-1 &   7.1$e$-2 \\ [0.1cm]
FF-SC      &   2.3$e$-5      & 7.8$e$-4       &  2.7$e$-5       &     8.9$e$-4           &  5.1$e$-3   &    8.6$e$-4       &   3.5$e$-3             &  1.1$e$-2   &   3.2$e$-3        &    7.3$e$-2      &   9.7$e$-2 &  6.9$e$-2 \\  \bottomrule
\end{tabular}
\end{table*}
\begin{table*}[ht!]
\vspace{-0.2cm}
\centering
\begin{tabular}{@{}ccccccccccccc@{}}
\toprule
\multirow{3}{*}{} & \multicolumn{12}{c}{$C=100, d_k=\{3,4,6\}$}                             \\ \cmidrule(l){2-13} 
Scenarios         & \multicolumn{3}{c}{offered load=100} & \multicolumn{3}{c}{offered load=150} & \multicolumn{3}{c}{offered load=200} & \multicolumn{3}{c}{offered load=250}\\ \cmidrule(l){2-13} 
                                      & Sim. & App.EES & App.SOC & Sim. & App.EES & App.SOC & Sim. & App.EES & App.SOC & Sim & App.EES & App.SOC \\ \cmidrule(l){1-13}      
RF            &    4.8$e$-3           &   \multirow{2}{*}{  2.8$e$-3} & \multirow{2}{*}{  1.4$e$-3}            &  3.6$e$-2             &   \multirow{2}{*}{  2.5$e$-2} & \multirow{2}{*}{  1.6$e$-2}           &   8.9$e$-2            & \multirow{2}{*}{  7.0$e$-2} & \multirow{2}{*}{  5.3$e$-2} &    1.5$e$-1           &   \multirow{2}{*}{  1.2$e$-1} &  \multirow{2}{*}{  1.1$e$-1}  \\ [0.1cm]
FF      &        5.6$e$-4           &   &            &  1.6$e$-2             &    &         &   6.3$e$-2            &   &   & 1.2$e$-1          &          &        \\ [0.1cm]
RF-SC            &    1.4$e$-3           &   \multirow{2}{*}{ 1.8$e$-3} & \multirow{2}{*}{ 5.3$e$-4}  &  1.5$e$-2   &   \multirow{2}{*}{ 1.9$e$-2} & \multirow{2}{*}{ 7.9$e$-3}           &   5.1$e$-2            & \multirow{2}{*}{ 5.9$e$-2} & \multirow{2}{*}{ 3.2$e$-2} & 1.0$e$-1           &   \multirow{2}{*}{ 1.1$e$-1} &  \multirow{2}{*}{ 7.2$e$-2}   \\ [0.1cm] 
FF-SC        &     3.1$e$-4           &   &          &   9.1$e$-3             &    &          &    3.9$e$-2      &     &    &  8.5$e$-2          &    & \\  \bottomrule 
 &\multicolumn{12}{c}{$C=200, d_k=\{4, 6, 10\}$}  \\ \midrule   
 RF       & 5.3$e$-4   & \multirow{2}{*}{ 1.9$e$-5}    &\multirow{2}{*}{ 8.8$e$-6}   &  9.3$e$-3 & \multirow{2}{*}{ 1.2$e$-3} &\multirow{2}{*}{ 8.5$e$-4}    & 3.6$e$-2 & \multirow{2}{*}{ 1.1$e$-2}  &\multirow{2}{*}{ 8.4$e$-3}   & 7.5$e$-2  & \multirow{2}{*}{ 3.5$e$-2} &\multirow{2}{*}{ 2.9$e$-2} \\ [0.1cm]
  FF       &  2.6$e$-6   &     &   &  7.8$e$-4 &   &   &  1.0$e$-2 &   &   &  3.8$e$-2  &  &  \\ [0.1cm]
RF-SC   &   1.1$e$-4   & \multirow{2}{*}{ 1.1$e$-5}  &\multirow{2}{*}{3.3$e$-6}   &   2.8$e$-3  & \multirow{2}{*}{ 9.7$e$-4}  &\multirow{2}{*}{ 3.9$e$-4}   &   1.5$e$-2  & \multirow{2}{*}{ 9.2$e$-3} &\multirow{2}{*}{ 4.5$e$-3}   &  3.8$e$-2 & \multirow{2}{*}{ 3.1$e$-2}  &\multirow{2}{*}{1.8$e$-2}  \\ [0.1cm]
FF-SC   &  2.5$e$-6   &   &    &   4.8$e$-4  &  &   &   6.2$e$-3  &  &   &   2.3$e$-2 &   &    \\ \bottomrule
\end{tabular}
\end{table*}

\subsection{Network Model}
\par Firstly, we consider a well known 14-node NSFNET topology with 42 unidirectional links and all possible OD pairs routes ($|\mathcal{O}|=182$) over which connection requests arrive according to a Poisson process. Table \ref{table:NSFresults} presents the BP results using $App.EES$ and $App.SOC$ and verifying simulations in  for a 14-node NSFNET  under various scenarios. Similar to a single-hop system, here $App.EES$ BPs are very close to Sim. BPs under the RF policy, and interestingly $App.SOC$ BPs are closer to Sim. BPs for the FF policy. On a closer look, we can observe that $App.SOC$ BPs under the RF policy is also very close to the Sim. BPs under the FF policy. This is very helpful in obtaining approximate BPs under the FF policies without the need to generate valid exact states for medium and large scale networks. For the medium and large-scale scenarios in Table \ref{table:NSFresults}, we see the similar trend, as observed in the link scenario, the $App.EES$ and $App.SOC$ BPs are close to Sim. BPs under the RF and the FF policies, respectively. However, again, in a large scale EON ($C=200$) $App.EES$ and $App.SOC$ BPs could differ with the simulation results for lower loads. When the network allows the SC operation at intermediate nodes, we observe that  BP reduces considerably under the RF-SC scenarios, as compare to the RF operation mode, i.e., without SC. The reason is that RF policy tries to assign random continuous and contiguous free slices to a new request, and with SC, the continuity constraint is relaxed when it does not find the required aligned free consecutive slice over a route.  However, the FF-SC operation still offers the lowest BPs as compare to other scenarios. Similar to the RF and the FF, the approximate BPs in RF-SC and the FF-SC  operations can also be obtained by $App.EES$ and $App.SOC$, respectively, and they also seem to be very close to the Sim. results for various loads and link  capacities.

\begin{table*}[ht!]
\centering
\caption{The BP in a 6-node ring network under various capacities and class of demands.} 
\label{table:RingResults}
\begin{tabular}{@{}ccccccccccccc@{}}
\toprule
\multirow{3}{*}{} & \multicolumn{12}{c}{$C=10, d_k=\{3,4\}$ }                             \\ \cmidrule(l){2-13} 

Scenarios        & \multicolumn{3}{c}{offered load=0.1} & \multicolumn{3}{c}{offered load=0.6} & \multicolumn{3}{c}{offered load=1.2} & \multicolumn{3}{c}{offered load=2.4}\\ \cmidrule(l){2-13} 
          & Sim. & App.EES & App.SOC & Sim. & App.EES & App.SOC & Sim. & App.EES & App.SOC & Sim & App.EES & App.SOC \\ \cmidrule(l){1-13}                                          
RF     &    1.0$e$-3         &    2.3$e$-3      &   2.2$e$-4         &     1.1$e$-2           &    1.8$e$-2 &  6.6$e$-3            &   3.0$e$-2             &   4.3$e$-2 &  2.3$e$-2           &    7.9$e$-2            &  1.0$e$-1 &  6.8$e$-2 \\ [0.1cm] 
FF      &     1.7$e$-4        &    3.5$e$-3       &    2.8$e$-4       &     5.5$e$-3           &  2.2$e$-2  &    7.1$e$-3          &   1.9$e$-2             &  4.7$e$-2   &    2.3$e$-2       &    5.8$e$-2            &  9.7$e$-2  &   6.4$e$-2\\  [0.1cm]
RF-SC  &     9.8$e$-4      &    1.3$e$-3     &     1.6$e$-4      &     1.0$e$-2           &    1.2$e$-2  &     5.2$e$-3          &   2.7$e$-2             &   3.3$e$-2 &     1.8$e$-2 &    7.2$e$-2            &  8.5$e$-2 &    5.7$e$-2 \\ [0.1cm]
FF-SC  &     1.7$e$-4      &   2.0$e$-3   &   1.8$e$-4   &     5.3$e$-3           &  1.5$e$-2   &     5.3$e$-3       &   1.8$e$-2             &  3.4$e$-2   &    1.8$e$-2        &    5.5$e$-2      &   7.9$e$-2 &   5.5$e$-2 \\  \bottomrule
\end{tabular}
\end{table*}


\begin{table*}[ht!]
\vspace{-0.2cm}
\centering
\begin{tabular}{@{}ccccccccccccc@{}}
\toprule
\multirow{3}{*}{} & \multicolumn{12}{c}{$C=100, d_k=\{3,4,6\}$ }                             \\ \cmidrule(l){2-13} 
Scenarios         & \multicolumn{3}{c}{offered load=50} & \multicolumn{3}{c}{offered load=100} & \multicolumn{3}{c}{offered load=150} & \multicolumn{3}{c}{offered load=200}\\ \cmidrule(l){2-13} 
                                      & Sim. & App.EES & App.SOC & Sim. & App.EES & App.SOC & Sim. & App.EES & App.SOC & Sim & App.EES & App.SOC \\ \cmidrule(l){1-13}      
RF            &    1.9$e$-2           &   \multirow{2}{*}{  1.3$e$-2} & \multirow{2}{*}{  6.9$e$-3}            &  1.4$e$-1             &   \multirow{2}{*}{  1.2$e$-1} & \multirow{2}{*}{  1.0$e$-1}           &   2.5$e$-1            & \multirow{2}{*}{  2.4$e$-1} & \multirow{2}{*}{  2.2$e$-1} &    3.4$e$-1           &   \multirow{2}{*}{  3.3$e$-1} &  \multirow{2}{*}{  3.1$e$-1}  \\ [0.1cm]
FF      &        5.8$e$-3           &   &            &  1.1$e$-1             &    &         &   2.3$e$-1            &   &   & 3.2$e$-1          &          &        \\ [0.1cm]
RF-SC            &      6.5$e$-3           &   \multirow{2}{*}{  9.4$e$-3} & \multirow{2}{*}{  3.0$e$-3}  &    1.0$e$-1   &   \multirow{2}{*}{  1.2$e$-1} & \multirow{2}{*}{ 7.7$e$-2}           &     2.2$e$-1            & \multirow{2}{*}{  2.3$e$-1} & \multirow{2}{*}{  1.9$e$-1} &   3.3$e$-1           &   \multirow{2}{*}{  3.3$e$-1} &  \multirow{2}{*}{  2.9$e$-1}   \\ [0.1cm] 
FF-SC       &       2.6$e$-3           &   &          &     8.9$e$-2             &    &          &      2.0$e$-1      &     &    &    3.1$e$-1          &    & \\  \bottomrule 
 &\multicolumn{12}{c}{$C=200, d_k=\{4, 6, 10\}$ }  \\ \midrule   
 RF       &  4.2$e$-3   & \multirow{2}{*}{  1.4$e$-4}    &\multirow{2}{*}{  7.4$e$-5}   &   7.4$e$-2 & \multirow{2}{*}{  4.1$e$-2} &\multirow{2}{*}{  3.5$e$-2}    &  1.6$e$-1 & \multirow{2}{*}{  1.3$e$-1}  &\multirow{2}{*}{  1.2$e$-1}   &  2.4$e$-1  & \multirow{2}{*}{  2.1$e$-1} &\multirow{2}{*}{  2.0$e$-1} \\ [0.1cm]
  FF       &   7.8$e$-5   &     &   &   4.2$e$-2 &   &   &   1.2$e$-1 &   &   &   2.0$e$-1  &  &  \\ [0.1cm]
RF-SC   &    6.3$e$-4   & \multirow{2}{*}{  9.7$e$-5}  &\multirow{2}{*}{  3.2$e$-5 }   &    4.6$e$-2  & \multirow{2}{*}{   3.8$e$-2}  &\multirow{2}{*}{   2.1$e$-2}   &    1.3$e$-1  & \multirow{2}{*}{   1.3$e$-1} &\multirow{2}{*}{   9.6$e$-2}   &   2.2$e$-1 & \multirow{2}{*}{   2.1$e$-1}  &\multirow{2}{*}{  1.8$e$-1}  \\ [0.1cm]
FF-SC   &   2.7$e$-5   &   &    &    2.6$e$-2  &  &   &    1.0$e$-1  &  &   &    1.8$e$-1 &   &    \\ \bottomrule
\end{tabular}
\end{table*} 

\par  Finally, we present approximate and verifying simulation BP results in Table \ref{table:RingResults} for a 6-node ring topology with 5 bidirectional links, and with all possible OD pairs routes ($|\mathcal{O}|=30$) over which connection path requests arrive according to a Poisson process. Since we route an OD pair request over its shortest path, each of 5 unidirectional (clockwise) links shares 6 OD routes, and each of other 5 unidirectional (anti-clockwise) links shares 3 OD pair routes. Thus,  as noted in \cite{sridharan2004blocking}, a ring topology as a sparse network is well suited for verifying the accuracy of approximate BP approaches.  We observe that $App.EES$ and $App.SOC$ BPs obtained under the RF and FF operations with and without SC, respectively are  acceptable, as they are closer to the simulation results under varying conditions, including link capacities, demands and traffic loads. SC operation is indeed useful in reducing blocking under the RF policy for lower and medium loads. Also, the similar trend is depicted at very low load in a large scale ring network, i.e., approximate BPs could be lower than the simulation results. However, we can not say for surety  whether approximate BPs obtained for both NSFNET and the ring network are underestimated or overestimated, due to different effects of the independence model and the reduced load approximation. Nevertheless, irrespective of the loads, classes of demands, and link capacity, $App.EES$ ($App.SOC$) can be used for obtaining BPs under the RF and RF-SC (FF and FF-SC) operations.

As both RF and FF policies have some advantages and disadvantages, e.g., FF is preferable for lower blocking, whereas RF is suitable for load balancing, security and lower level of crosstalk in space-division-multiplexing-enabled EONs \cite{fujii2014demand,singh2017combined}. Thus both policies can be made useful for deployment of new services. In summary, we can say that $App.EES$ can be used by network operator to estimate BP in EONs with or without SC under the RF policy, and $App.SOC$ for the FF policy.  Nevertheless, the accuracy of approximate blocking probability could be further improved by utilizing a more accurate probability of acceptance of a request in EONs without SC, and considering the link correlation model used in WDM networks \cite{sridharan2004blocking}, which is relatively complex compared to the link independence model but essential to analyze the effect of correlation of loads among links under different spectrum allocation policies in a network. At the same time, the scalability of load and link correlation models could be a major issue that needs to be worked on in the future. 

\section{Conclusions} \label{sec:conclusion}  
\par In this paper,  we proposed the first exact Markov model for analyzing blocking probability in EONs, and subsequently the related methods to reducing the exact link state occupancy model into a reduced occupancy model to computing approximate blocking.  More in detail, we presented load-independent and load-dependent approximations to compute the probability of acceptance of a request in EONs with and without spectrum conversion, considering bandwidth demands, contiguity constraint and continuity constraint. These approximations use the information of the number of non-blocking and blocking exact states corresponding to an occupancy state, which we derive for a random-fit assignment policy using inclusion-exclusion principle. Additionally, approximate BPs are presented for cases with and without spectrum conversion under random-fit and first-fit spectrum allocation policies. The numerical results obtained show that the exact blocking analysis is accurate, albeit limited to a very small scale EONs, due to complexity. On the other hand, approximate solutions have been shown accurate in a broader range of scenarios. It was shown in fact that the accuracy of the approximation methods proposed depend on the various factors, such as the spectrum allocation policies, link capacity, traffic loads, and topology. The next steps in this line of work include major challenges that have not been solved yet, and are analytically rather complex, most notably the interdependency of network link loads. 

\appendices 
\section{Derivation of total, Non-blocking and Blocking Exact Link States Under the RF Policy} \label{sec:Appendix1}
\par Without loss of generality, let us assume that there are $\mathrm{r}$ number of routes $o=1,\ldots, \mathrm{r}$ traversing a link under consideration. For a given microstate $x, 0\leq x\leq C$, where the number of empty (free) slices $E(x)=C-x$, we can find connection patterns or macrostates (i.e., connections per class per route) $\textbf{n}=(n_1^1,\ldots,, n_K^1,\ldots, n_1^\mathrm{r},\ldots, n_K^\mathrm{r})$ that satisfies $\textbf{n}\cdot\textbf{d}_\mathrm{r}^T=x$, where  $T$ is transpose, and $\textbf{d}_\mathrm{r}=(d_1^1,\ldots,d_K^1,\ldots,d_1^\mathrm{r},\ldots,d_K^\mathrm{r})$ is an array with $r \times K$ elements, and by definition for all routes $o$, $d_k^o=d_k$. Now, for each connection pattern $\textbf{n}=(n_1^1,\ldots, n_K^1,\ldots, n_1^\mathrm{r},\ldots, n_K^\mathrm{r})$, the $E(x)$ empty slices can be distributed at $N(\textbf{n})+1$ places (including the start, end, and in between each two connections), where the total number of connections $N(\textbf{n})= \sum_{k=1}^K\sum_{o=1}^\mathrm{r} n_k^o(\textbf{n})$. Noting that there are $\frac{N(\textbf{n})!}{\prod_{k=1}^K\prod_{o=1}^rn_k^o(\textbf{n})!}$ distinct permutations of connections in $\textbf{n}$,  and there are $\binom{E(x)+N(\textbf{n})}{N(\textbf{n})}$ different ways to distribute $E(x)$ empty slices at $N(\textbf{n})+1$ places in each unique permutation of $\textbf{n}$, the number of exact states with all connection patterns representing a microstate occupancy $x$ is, thus, given by 
\begin{equation} 
|\Omega_S(x)| = \sum_{\textbf{n} \in \Omega_S(x)}\frac{N(\textbf{n})!}{\prod_{k=1}^K\prod_{o=1}^\mathrm{r}n_k^o(\textbf{n})!}\times\binom{E(x)+N(\textbf{n})}{N(\textbf{n})}. \nonumber
\end{equation}

Importantly, only some of the  exact link states ($\textbf{s} \in \Omega_S(x) $) are non-blocking states, as defined in Eq. \eqref{eqn:NB states}. To compute the number of non-blocking exact link states, let us solve the following equation for each permutation of the connection pattern  for all $\textbf{n} \in \Omega_S(x)$: 
\begin{equation}
a_1 + a_2 +\cdots + a_{N(\textbf{n})+1} = E(x), s.t., \exists i: a_i \geq d_k \nonumber
\end{equation}

Using the inclusion-exclusion principle (hint: consider an event $A_i= \{a_i\geq d_k\}, 1\leq i \leq N(\textbf{n})+1$ and find $|\cup_{i}A_i|$), the number of non-blocking exact states corresponding to each permutation of connections in $\textbf{n} \in \Omega_S(x)$ can be given by
\begin{equation}
W(\textbf{n}) =  \sum_{i=1}^{N(\textbf{n})+1} (-1)^{i+1} \binom{N(\textbf{n})+1}{i} \binom{E(x)+N(\textbf{n})-id_k}{N(\textbf{n})}. \nonumber
\end{equation}

Now, considering all permutations of $\textbf{n}$, and adding all non-blocking states corresponding to each $\textbf{n}$ belonging to the microstate $x$ would result in the number of class $k$ non-blocking exact states for a given microstate $x$, given by  
\begin{equation} 
|\mathbb{NB}(x,k)|= \sum_{ \textbf{n} \in \Omega_S(x)}W(\textbf{n})\times \frac{N(\textbf{n})!}{\prod_{k=1}^K\prod_{o=1}^\mathrm{r}n_k^o(\textbf{n})!}. \nonumber
\end{equation}
Noting that for a class $k$ request in any occupancy state $x$, $|\Omega_S(x)|= |\mathbb{NB}(x,k)|+|\mathbb{FB}(x,k)|+|\mathbb{RB}(x,k)|$, the number of fragmentation blocking exact states in a microstate $x, 0\leq x \leq C-d_k$ is $|\mathbb{FB}(x,k)|=|\Omega_S(x)|-|\mathbb{NB}(x,k)|$, since $|\mathbb{RB}(x,k)|=0$. On the other hand, all exact states representing a microstate $x, C-d_k < x \leq C$  are resource blocking states for class $k$ request, i.e., $|\mathbb{RB}(x,k)|=|\Omega_S(x)|$, and the number of non-blocking and fragmentation blocking exact states are both zero.

\section{Uniform Approximation for Computing Probability of Acceptance} \label{sec:Appendix2}
In this Appendix, we derive an analytical expression for computing approximate probability of acceptance $p_k(\textbf{x}_r)\equiv p_k(x_{j_1},x_{j_2},\ldots, x_{j_l})= Pr[Z_r\geq d_k|X_{j}=x_{j_1},\ldots, X_{j_l}=x_{j_l}]$ on an $l$-hop route in an EON using a \textit{Uniform} approach. Considering the independence link assumption, we further assume that  \emph{the occupancy of slices are independent and identically distributed in each link.} This means that total occupied slices are uniformly distributed, i.e., the spectrum patterns are formed by a single slice-demand with a given total occupancy, without considering the contiguous allocation of slices, and are not restricted to the spectrum patterns generated by a  given classes of demands and spectrum allocation policy. Now, for a given occupancy of links on a route $r$, the probability that there are $n$ continuous (but not necessarily contiguous) free slices on its route is obtained by the following recursive relationship \cite{birman1996computing,kuppuswamy2009analytic}:
\begin{align}
 g_n(x_{j_1},x_{j_2},\cdots, x_{j_l}) = Pr[Z_r = n|X_{j_1}=x_{j_1},\cdots, X_{j_l}=x_{j_l}] & \nonumber \\ 
 = \sum_{i=n}^{i^*}g_n(C-i, x_{j_l}) g_i(x_{j_1}, x_{j_2},\cdots, x_{j_{l-1}}) &
\end{align}
where $i^*=min(C- x_{j_1}, C-x_{j_2},\cdots, C-x_{j_{l-1}})$ and $g_n(x, y)= \binom{C-x}{n}\binom{x}{C-y-n}/\binom{C}{C-y}$. 


Now, we could find the probability that the route $r$ has equal or more than $d_k$ free contiguous slices  $\{Z_r\geq d_k\}$ across links on its route, given the link occupancy vector $\textbf{x}_r$ and also there are exactly $n$ continuous free slices on the route with $\{X_r=n\}$, i.e., $p_k^{Uni.}(\textbf{x}_r, n)= Pr[Z_r\geq d_k|X_{j_1}=x_{j_1},\cdots, X_{j_l}=x_{j_l}, X_r=n]$. Using the inclusion-exclusion principle, it can be given by 
\begin{equation}
p_k^{Uni.}(\textbf{x}_r, n) = \frac{\sum_{i=1}^{C-n+1} (-1)^{i+1} \binom{C-n+1}{i}\binom{C-id_k}{C-n}}{\binom{C}{n}}.
\end{equation} 
The above equation seems to be independent of $\textbf{x}_r$, but actually a factor which is a function of $\textbf{x}_r$ is multiplied in both numerator and denominator, thus cancels the effect of $\textbf{x}_r$. 
Thus, $p_k^{Uni.}(\textbf{x}_r) = Pr[Z_r\geq d_k|X_{j_1}=x_{j_1},\cdots, X_{j_l}=x_{j_l}]$ can be given as follows.
\begin{equation}
p_k^{Uni.}(\textbf{x}_r) =\sum_{n=d_k}^{min(C- \textbf{x}_r)} p_k(\textbf{x}_r, n)g_n(\textbf{x}_r)
\label{eqn:p_k_x_uniform}
\end{equation}

Under the \textit{Uniform} approximation, the probability of acceptance in EONs with SC can easily be given by using Eq. \eqref{eqn:ApprxProbOfAcceptInRoute-SC}. Noting that the spectrum patterns are assumed to be created by a single slice demand in the \textit{Uniform} approximation. Thus, the probability that a link $j$ in state $x_{j}$ have equal or more than $d_k$ free consecutive slices can be given by the ratio of non-blocking and total exact states in $x_{j}$ as follows, which uses $\textbf{n}=(n_1)=(x_j)$ and $N(\textbf{n})=x_{j}$ in Eqs. \eqref{eqn:AllNBstates4microstate} and  \eqref{eqn:Allstates4microstate}. 
\begin{equation}
p_{k,sc}^{Uni.}(x_{j})  =  \frac{\sum_{i=1}^{x_{j}+1} (-1)^{i+1} \binom{x_{j}+1}{i} \binom{C-id_k}{x_{j}}}{\binom{C}{x_{j}}}
\end{equation}
Finally, the probability of acceptance of a request with demand $d_k$ on a route $r(o)$ with $l$-hops in an EON with SC can be obtain by multiplying link acceptance probabilities ($p_{k,sc}^{Uni.}(x_{j})$) on the route $r(o)$, as shown by Eq. \eqref{eqn:ApprxProbOfAcceptInRoute-SC}.

\bibliographystyle{IEEEtran}
\bibliography{BPbib}

\end{document}